\documentclass[sigconf,screen]{acmart}
\AtBeginDocument{%
  \providecommand\BibTeX{{%
    \normalfont B\kern-0.5em{\scshape i\kern-0.25em b}\kern-0.8em\TeX}}}

\setcopyright{rightsretained}
\renewcommand\footnotetextcopyrightpermission[1]{}

\acmPrice{15.00}
\usepackage{graphicx}
\usepackage{svg}
\usepackage{wrapfig}
\usepackage{paralist}
\usepackage{caption}
\usepackage{subcaption}
\usepackage{cleveref}
\usepackage{paralist}
\usepackage{enumitem}
\usepackage{microtype}
\DisableLigatures{encoding = *, family = * }

\usepackage{multirow}
\usepackage{pifont}

\usepackage{colortbl}

\definecolor{tableheader}{HTML}{EFEFEF}
\definecolor{tablegrayline}{HTML}{a0a0a0}

\usepackage{listings}
\usepackage{xcolor}
\usepackage{multirow}
\usepackage{adjustbox}
\usepackage{hhline}

\newcommand{\xmark}{\ding{55}}%

\usepackage{tikz}

\lstdefinestyle{mystyle}{
  backgroundcolor=\color{backcolour}, commentstyle=\color{codegreen},
  keywordstyle=\color{magenta},
  numberstyle=\tiny\color{codegray},
  stringstyle=\color{codepurple},
  basicstyle=\ttfamily\footnotesize,
  breakatwhitespace=false,         
  breaklines=true,                 
  captionpos=b,                    
  keepspaces=true,                 
  numbers=left,                    
  numbersep=5pt,                  
  showspaces=false,                
  showstringspaces=false,
  showtabs=false,                  
  tabsize=2
}

\lstdefinestyle{General} {
    basicstyle=\small\ttfamily,
    breaklines=true
}
\definecolor{codegreen}{rgb}{0,0.6,0}
\definecolor{codegray}{rgb}{0.5,0.5,0.5}
\definecolor{codepurple}{rgb}{0.58,0,0.82}
\definecolor{backcolour}{rgb}{0.95,0.95,0.92}
\newcommand{\eg}{\textit{e.g.,\ }}

\newcommand{\etal}{\textit{et al.\ }}
\newcommand{\etals}{\textit{et al.'s\ }}

\newcommand{\vs}{\textit{vs\ }}

\usepackage{comment}
\usepackage{ifthen}
\usepackage{xspace}
\usepackage{etoolbox}
\newtoggle{comments}
\toggletrue{comments}

\usepackage{hyperref}

\newcommand{\initialNumberOfNotebooks}{10000
}

\newcommand{\numberOfNotebooks}{100000
}

\newcommand{\notebookScaleFactor}{10x
}

\newcommand{\overlapInDatasets}{87
}

\newcommand{\validFileCountAfterNBConvert}{99441
}

\newcommand{\filteredFileCountDuringNBConvert}{559
}

\newcommand{\numberOfPythonNotebooks}{92050
}

\newcommand{\numberOfNotebooksWithNoLanguageInfo}{2672
}

\newcommand{\numberOfNotebooksWithSourceOutputs}{39540}

\newcommand{\percentageOfNotebooksWithSourceOutputs}{42.95\%
}

\newcommand{\numberOfNotebooksWithCodeOnly}{52509 (57.04\%)
}

\newcommand{\totalNumberOfHTMLFilesAcrossThemes}{589746
}

\newcommand{\numberOfHTMLFilesPerTheme}{98291
}

\newcommand{\totalNumberOfImages}{342722}

\newcommand{\totalNumberOfImagesWithoutAlt}{342102 (99.81\%)}

\newcommand{\totalNumberOfImagesWithAlt}{609
}

\newcommand{\totalNumberOfImagesWithAltFromCode}{1
}

\newcommand{\totalNumberOfImagesWithAltFromMarkdown}{608
}

\newcommand{\medianImagesPerNotebookFiguresOnly}{4.0
}

\newcommand{\pNinetyNumberOfImages}{16
}

\newcommand{\pNinetyNineNumberOfImages}{77
}

\newcommand{\maxNumberOfFiguresInNotebook}{12858
}

\newcommand{\percentageNotebooksWithNoTables}{65.9\%
}

\newcommand{\percentageNotebooksWithTables}{34.1\%
}

\newcommand{\meanNumberOfTablesInNotebooks}{5.55
}

\newcommand{\medianNumberOfTablesInNotebooks}{3.0
}

\newcommand{\maximumNumberOfTables}{1181
}

\newcommand{\lastPercentileNumberOfTables}{37.0
}

\newcommand{\numberOfApplicationMimeTypes}{24
}

\newcommand{\numberOfTypesOfTextTypes}{6
}

\newcommand{\numberOfTypesOfImageTypes}{4
}

\newcommand{\topFiveOutputsCumulativePercentage}{98.67\%
}

\newcommand{\numberOfMedianRowsInTables}{6
}

\newcommand{\numberOfMedianColumnsInTables}{30
}

\newcommand{\numberOfMeanRowsInTables}{15
}

\newcommand{\numberOfMeanColumnsInTables}{140
}

\newcommand{\largestTableContainsRows}{162736
}

\newcommand{\largestTableContainsColumns}{1139145
}

\newcommand{\largestTableMaxCellCount}{185379900720
}

\newcommand{\totalNumberOfScanResults}{238675580
}

\newcommand{\totalCellsWithOutputGraphicsFromCodeCells}{208427
}

\newcommand{\totalCellsWithOutputGraphicsAndNoHeadingAfter}{159341
}

\newcommand{\totalCellsWithOutputGraphicsAndNoHeadingAfterContainingMarkdownBeforeOrAfter}{59166
}

\newcommand{\totalCellsWithOutputGraphicsAndNoHeadingAfterContainingTablesBeforeOrAfter}{13037
}

\newcommand{\totalCellsWithOutputGraphicsAndNoHeadingAfterContainingMarkdownAndContainsTablesBeforeOrAfter}{2566
}

\newcommand{\numberOfNotebooksWithPossibleManualTablesAndSupportingExplanation}{1795
}

\newcommand{\numberOfNotebooksWithPotentialRelatedTables}{7109
}
\newcommand{\numberOfNotebooksWithTables}{33517
}

\newcommand{\numberOfNotebooksWithNoTables}{64774
}

\usepackage{fancyhdr}
\usepackage[framemethod=tikz]{mdframed}

\copyrightyear{2023}
\acmYear{2023}
\setcopyright{rightsretained}
\acmConference[ASSETS '23]{The 25th International ACM SIGACCESS Conference on Computers and Accessibility}{October 22--25, 2023}{New York, NY, USA}
\acmBooktitle{The 25th International ACM SIGACCESS Conference on Computers and Accessibility (ASSETS '23), October 22--25, 2023, New York, NY, USA}
\acmDOI{10.1145/3597638.3608417}
\acmISBN{979-8-4007-0220-4/23/10}

\begin{document}

\pagestyle{fancy}
\fancyhf{}
\fancypagestyle{firststyle}
{
   \fancyhf{}
   \fancyhead[C]{
   \begin{mdframed}
   \centering
       This is a pre-print of the paper accepted to ACM ASSETS'23. A screen reader accessible PDF version is available \href{https://blvi.dev/noteably-inaccessible-paper}{for download here}.
   \end{mdframed}
   }
   \renewcommand{\headrulewidth}{0pt} % removes horizontal header line
}

\title[Data Science Notebook (In)Accessibility]{Notably Inaccessible -- Data Driven Understanding of Data Science Notebook (In)Accessibility}


\settopmatter{authorsperrow=2}
\author{Venkatesh Potluri}
\authornote{Equal contributors}
\orcid{0000-0002-5027-8831}
\email{vpotluri@cs.washington.edu}
\affiliation{
\country{}
}
\author{Sudheesh Singanamalla}
\authornotemark[1]
\orcid{0000-0003-2462-7273}
\email{sudheesh@cs.washington.edu}
\affiliation{
\country{}
}
\author{Nussara Tieanklin}
\orcid{0000-0002-3074-8699}
\email{nussara@cs.washington.edu}
\affiliation{
\country{}
}
\author{Jennifer Mankoff}
\orcid{0000-0001-9235-5324}
\email{jmankoff@cs.washington.edu}
\affiliation{
\country{}
}
\author{}
\affiliation{%
   \institution{University of Washington}
   \department{Paul G. Allen School of Computer Science and Engineering}
   \city{Seattle}
   \state{WA}
   \country{USA}}

\renewcommand{\shortauthors}{Potluri et al.}
\renewcommand{\authors}{Venkatesh Potluri, Sudheesh Singanamalla, Nussara Tieanklin, Jennifer Mankoff}

% \input{sections/meta/ccs-keyword}



\begin{abstract}
	Computational notebooks, tools that facilitate storytelling through exploration, data analysis, and information visualization, have become the widely accepted standard in the data science community. These notebooks have been widely adopted through 
	notebook software such as Jupyter, Datalore and Google Colab, both in academia and industry.  While there is extensive research to learn  how data scientists use computational notebooks, identify their pain points, and enable collaborative data science practices, very little is known about the various accessibility barriers experienced by blind and visually impaired (BVI) users using these notebooks. BVI users are unable to use computational notebook interfaces due to
	\begin{inparaenum}
		\item inaccessibility of the interface,
		\item common ways in which data is represented in these interfaces, and
		\item inability for popular libraries to provide accessible outputs.
	\end{inparaenum}
	We perform a large scale systematic analysis of 100000 Jupyter notebooks to identify various accessibility challenges in published notebooks affecting the creation and consumption of these notebooks. Through our findings, we make recommendations to improve accessibility of the artifacts of a notebook, suggest authoring practices, and propose changes to infrastructure to make notebooks accesible. \textcolor{blue}{An accessible PDF can be obtained at \url{https://blvi.dev/noteably-inaccessible-paper}}
\end{abstract}

%\input{sections/meta/css-keyword.tex}


\maketitle
\thispagestyle{firststyle}

% \blfootnote{$\star$ Equal Contribution}

\section{Introduction}
\label{section:introduction}

Computational notebooks such as Jupyter~\cite{jupyter} combine code, natural language, and rich representations of data providing a ubiquitous literate programming experience~\cite{knuth1984literate}. These notebooks are used through computational notebook systems and programming environments such as Jupyter, Google Colab, Datalore, and Noteable (among others) that abstract software setup, computational infrastructure management and access to resources. Computational notebooks are widely used by data scientists  as interactive mechanisms to process, understand and express data making it easier for them to collaborate, share code, convey stories and narratives through data visualizations and text, while keeping the reproducibility of results in mind~\cite{wang2020better, rule2018exploration}. 
The popularity of these computational notebooks, specifically Jupyter notebooks, as the go-to tool for data science can be seen in the rapid increase in published notebooks, 2.5 Million public notebooks hosted on GitHub in September 2018~\cite{rule2018exploration}, increasing by 10x since 2015~\cite{wang2020better}. As of 2020, this dataset has increased to over 10 Million and has been analyzed by JetBrains -- a company building Integrated Development Environments (IDEs)~\cite{guzharina2020jetbrains}.

Despite computational notebooks being popular tools, we know very little about the accessibility of these tools for developers and data scientists who are blind or visually impaired (BVI). 
What little has been written on the topic is found in non-peer reviewed sources such as  Astronomy Notebooks for All~\cite{notebooks4all}, an effort to perform accessibility auditing of the Jupyter Lab interface and contribute  changes to the upstream Jupyter open source community. An early 2023 analysis of the accessibility score for Jupyter Hub  graded it as a fail (F)~\cite{manfromjupyter2020accessibility}. One active effort to address the inaccessibility of notebook software can be found in Microsoft's VisualStudio Code, a popular IDE among the BVI developer community that is building a new, more accessible notebook authoring experience on top of the existing standardized format of Notebooks by adding improved keyboard navigation and audio cues.
While these are much needed improvements to the Notebook IDE experience, they do not contribute to our understanding of the full variety of accessibility issues that can arise from  from the different ways in which computational notebooks are authored, consumed, and published. Understanding these accessibility issues can be critical to improving the accessibility of notebooks and the infrastructure that supports their creation, consumption, and distribution.

We present a data-driven investigation of the accessibility of computational notebooks. Our investigation focuses on accessibility for BVI notebook \textit{authors} and \textit{consumers} (hereby referred to as BVI users or BVI notebook users). 
Our work answers the question of whether IDE artifacts, authoring experiences, and infrastructure to work with computational notebooks, are accessible to BVI users. We answer the following specific research questions:
\begin{enumerate}[label=\textit{\textbf{RQ\arabic*}}]
    \item \textit{Data Artifacts}: How accessible are key data artifacts, namely \textit{figures and tables}, to blind or visually impaired users? 
    \item \textit{Authoring}: How do existing notebook authoring practices impact screen reader users' ability to glance through important information and results? For example, do most notebooks make proper use of headers and other landmarks that improve navigation and glanceability?
    \item \textit{Infrastructure}: How do current tools to distribute and customize notebooks impact accessibility? For example, how do different themes  for coloring a notebook impact the number of accessibility errors found by automated tools assessing that notebook?
\end{enumerate}

We answer these questions in the context of a large-scale, in-the-wild dataset of computational notebooks. This includes notebooks that may have been used for anything from exploratory data analysis to documents produced for public distribution . These notebooks could be written and consumed by users from a variety of disciplines including students, coders, or data scientists, and this process should be accessible to any notebook user who may be blind or visually impaired at any stage of the notebook authoring or consumption process. Thus, we chose to assess accessibility at scale without narrowing to a specific use category.

We narrow our our analysis to a subset of 100,000 notebooks selected at random from the JetBrains dataset of 10,000,000 (10 Million) of them~\cite{guzharina2020jetbrains}. Choosing such a large, random sample helps us to understand common patterns in notebook inaccessibility.  We complement this with manual verification of  a smaller set of 10  notebooks. Additionally, we narrow parts of our analysis to Python, the most popular language used in notebooks, so that we can perform language-specific code analysis to gain a deeper understanding of inaccessibility caused by notebook infrastructure. 

Our contributions are as follows: 
\begin{enumerate}
    \item We develop repeatable automated metrics that represent \textit{optimistic upper bounds} for estimating notebook accessibility.
    \item We developed a method for the first systematic large scale analysis of the current state of accessibility of computational notebooks to blind or visually impaired notebook authors / consumers. We open source our dataset and processing pipeline to enable researchers to build on this method and extend our results~\cite{potluri_venkatesh_2023_8185050}.
    \item We present results highlighting the overall inaccessibility of notebooks with respect to three research questions which look at the accessibility of data artifacts, notebook IDEs, and infrastructure. We also describe the programming tools most commonly used by notebook authors.
    \item Based on our results,  we highlight opportunities such as  encouraging good ALT text authoring practices, or automatically generating an accessible table alongside a chart. We make recommendations for notebook software developers, data scientists, and accessibility researchers creating computational notebooks, to make data and the corresponding story telling accessible through computational notebooks.
\end{enumerate}

Our findings about accessibility of computational notebooks through this characterization can have the potential to quicken the pace of making data science accessible by identifying the right improvements to widely used tools and notebook authoring practices, reducing the need for bespoke, custom accessibility-only solutions.

\section{Background}
\label{ssec:background}

Computational notebooks have risen in popularity since the inception of Jupyter in 2014, impacting many domains within and outside of computer science such as data science, machine learning, and astronomy.  This impact has been recognized by the Association of Computing Machinery (ACM) in 2017 with a prestigious software systems award~\cite{acm-software-system-award}.

Often, these notebooks are authored in a web-based IDE such as Jupyter Lab and Jupyter book, or through hosted and managed alternatives such as Google Colab, and Datalore.  Since their invention, millions of such notebooks have been authored for data analysis and related tasks, and it is important that we analyze this phenomena~\cite{wang2020better} in the context of accessibility. Rule \etal{} collected and released a dataset of one million Jupyter notebooks~\cite{rule2018exploration}. Analyses of these notebooks have shown that the majority of the notebooks do not declare dependencies, and have not been tested ~\cite{pimentel2021understanding};  and  notebook users consider them to be personal and messy~\cite{rule2018exploration}. 

Although many notebooks are used primarily by their authors, some notebooks are \textit{published}. Publishing a notebook leverages tools built into the notebook IDE to generate a webpage, or sometimes a PDF or \LaTeX\ document. In the case of web pages, these tools use web semantics such as headings and tables to structure content, and allow notebooks to be themed or otherwise decorated.

To better understand what issues may be of concern, we review the literature outside of computational notebooks in three closely related areas: Programming / IDE accessibility (relevant to the accessibility of the notebook \textit{authoring} experience, \Cref{background:programming_ide_accessibility}); Data analysis and visualization accessibility (relevant to both \textit{authors} and \textit{consumers}, \Cref{background:programming_ide_accessibility}); and Web Accessibility (relevant to \textit{consumers} and authors of published notebooks \Cref{background:accessibility_of_web}).

\subsection{Programming/IDE Accessibility}
\label{background:programming_ide_accessibility}

Accessibility of web-based programming has been studied in the past in the context of High-fidelity prototyping tools, which were found to have inaccessible graphical user interface (GUI) controls preventing BVI users of these tools from accessing the content in widgets and manipulating them on the prototyping canvas~\cite{LiTS21}.

Another useful point of comparison consider accessibility concerns raised in studies of other (non-web) programming.
Mealin and Murphy-Hill published one of the first studies to understand accessibility barriers experienced by BVI developers~\cite{Mealin:2012:BVIDevExploratory}. 
They found challenges associated with using IDEs and developing user interfaces. Other studies have found that accessibility barriers can impact  navigation, debugging, and glanceability of  code during the programming process~\cite{Albusays:2017:IOBSDW,Potluri:2018:CT}. Accessibility can also impact other tasks related to software development such as information seeking and collaboration with sighted users~\cite{storer-chi21,pandey2021understanding,potluri:2022:CodeWalk}. 

Solutions to these accessibility concerns include improvements to IDEs, bespoke software tools, and physical tactile interfaces to make web and user interface development accessible to BVI developers~\cite{Li:2019:ESL, Potluri:AMA:2019, Shetty:2020:TWDBVIP, Li:2022:TangibleGrid, Pandey:2022:UIA11y}.
Several tools present novel, accessible representations of code by repurposing familiar navigational structures such as list views, tree views and tables  to facilitate efficient screen reader navigation~\cite{Baker:2015:SJ, Schanzer:2019:APBVI, Potluri:2018:CT, Md:2022:Grid-Coding}. 
Additionally, audio has been used as a feedback mechanism to facilitate accessible debugging of code~\cite{stefik2009sodbeans, Stefik:2007:WAD, Potluri:2018:CT}. 
However, the impact of these tools to  accessibly support data intensive programming has not been evaluated. 

\subsection{Accessibility of Data}
\label{background:accessibility_of_data}

A critical aspect of using computational notebooks is to create, consume, and collaborate on visual, tabular, and other representations of data. These representations are often generated as results of computations performed in a Jupyter notebook. Understanding prior work on data and accessibility will help contextualize accessibility barriers that prevent notebook users from accessing data and results of computations performed in these notebooks.

Several efforts have explored making data visualizations \textit{accessible} through auditory representations combining speech with tones, and \textit{interactive} through voice commands, keyboard shortcuts, and touch screen gestures~\cite{SharifCWR21,Holloway:2022:Infosonics, Sharif:2022:VoxLens,zong2022rich, appleaudiographs}. Sharif \etal{} present a JavaScript plugin that makes two-dimensional data accessible to screen reader users~\cite{Sharif:2022:VoxLens}. The plugin supports both speech based summaries  and sonifications to convey trends in the data, and can be used along with a screen reader. Zong \etal{} focus specifically on screen reader interactions of charts and inform that improvements to structural organization, navigation, and descriptions are necessary to improved screen readability of visualizations~\cite{zong2022rich}. Though not sufficient to make data visualizations accessible, they find that accompanying visualizations with tables is crucial for accessibility due to the familiarity of tables to screen reader users.

These efforts assume BVI people as non-expert  consumers of data visualizations.  Very few efforts have resulted in tools and interfaces for BVI people to author accessible data visualizations. As of today, the work by Cantrell \etal{} resulting in the development of Highcharts Sonification Studio is the only open source charting library to support accessible authoring of data sonifications~\cite{cantrell2021highcharts}. Potluri \etal{} developed a data sonification toolkit, centered around the needs of BVI developers' attempts at and need for understanding sensor data and enable them to develop Internet of Things (IoT) applications~\cite{Potluri:2022:PSST}.
Computational notebooks, in addition to enabling consumption, have the potential to give BVI developers the means to produce data visualizations and enable data driven story telling, therefore surfacing the need for these tools to offer capabilities to provide accessible data visualizations. The very few attempts to make data representations accessible to BVI computational notebook users resulted in libraries with very limited functionality, leaving much to be discovered about their accessibility. 

\subsection{Accessibility of Published Information}
\label{background:accessibility_of_web}

As mentioned earlier, many notebooks are published, and one common format for this is to turn them into a web page. Additionally, many notebook software use web interfaces and leverage web semantics such as headings and HTML tables to structure content and outputs.
Thus, coupling the findings from a domain specific tool leveraging web as a platform like Jupyter, with web accessibility helps us identify potential accessibility concerns that may be unique to notebooks. Studies have been conducted in the past that use web accessibility guidelines to examine and improve the accessibility of other domains. For example, Elavski \etal~\cite{Elavsky:2022:Chartability} extend web accessibility guidelines to make data visualizations accessible. Similarly, Li \etal use IBM's accessibility checklist -- a set of accessibility guidelines derived from web accessibility guidelines, to compensate for the lack of industry standards to examine the accessibility of high-fidelity prototyping tools~\cite{LiTS21}.
Our findings from the web accessibility analysis of Jupyter notebooks contribute to this body of work.

\subsubsection*{Summary of concerns relating to Notebook accessibility}

Our literature review highlights several areas in which accessibility problems might arise, including: Visualization and data table access, both of which come up frequently in notebooks; and navigation and glanceability during coding, which would  be relevant to the notebook authoring experience. We now describe our analysis pipeline to understand accessibility concerns with notebooks in the areas highlighted in prior literature and present our results.

\section{Studying Notebook Accessibility}
\label{section:analysis}

Our background section highlighted some important areas of relevance to examine, to identify accessibility  concerns that might impact the experience of notebook consumers and authors. Building on these observations, our study focuses on an at-scale assessment of the accessibility of the consumer and authoring experiences. We choose this approach because of its high ecological validity: while a user study must narrow their scope to a very small set of notebooks, possibly hand curated to be semi-accessible so that the study is not a waste of participants’ time, a large scale study can explore a much broader range of notebooks. Below we introduce our study approach and sampling strategy. We also discuss the metadata that we extract from notebooks to prepare for our analysis of results.

While there are several datasets of Jupyter notebooks available~\cite{rule2018exploration, quaranta2021kgtorrent},  with metadata about where they come from and how they are created, they do not fully capture the variety of contexts that computational notebooks could be used in. Further, accessibility issues can occur in notebooks irrespective of context, and tools should inherently support the creation, consumption, and distribution of accessible notebooks.

We begin by  detailing our data processing pipeline, including our approach to sampling. We  then explain our method of measuring accessibility. We defined our accessibility metrics to prioritize an optimistic upper bound (meaning metrics that have high sensitivity but low precision), because there does not currently exist a validated measure of automatically measuring true accessibility. As we will see, this approach ultimately allows us to confidently say that most in-the-wild notebooks are inaccessible (hinting at the fact that the situation is potentially even worse than we estimate). Put differently, our approach has lower \textit{construct validity} than a user study might have. 
To complement this automated measurement and analysis of accessibility, we also conducted experiments to manually verify notebook glanceability through screen reader testing.

\subsection{Data Sampling and Filtering}
\label{subsec:data_processing_pipeline}

We start with the dataset provided by JetBrains that contains 10 million Jupyter notebooks~\cite{guzharina2020jetbrains}. Because of the computational and time costs of analyzing a data set of this size,  we began by analyzing a sample of \initialNumberOfNotebooks randomly chosen notebooks from the 10 Million notebook dataset. We start with this random sample of \initialNumberOfNotebooks notebooks to test our analysis pipelines, and gain an understanding of the characterization of the dataset. After establishing the required analysis pipelines, we scaled our analysis by \notebookScaleFactor and obtained a new random subset of notebooks resulting in \numberOfNotebooks notebooks. We observed that the results from our analysis pipeline returned similar observations in both the \initialNumberOfNotebooks and \numberOfNotebooks notebook analysis, giving us the confidence that \numberOfNotebooks was a sufficient sample size to draw conclusions from. Therefore, we stopped our analysis without further scaling the number of notebooks in our dataset. By chance, there was  overlap of \overlapInDatasets notebooks between these two data sets, which we considered small enough to be inconsequential and retained all \overlapInDatasets overlapping notebooks  for our analysis.

It is likely that some of these notebooks may be intended for scratch use, or exploratory data analysis which might be inaccessible compared to the presentation-ready notebooks. Since our work intends to explore accessibility for BVI authors, as well as consumers, including these notebooks in our study is intentional. Only studying notebooks that are presentation-ready assumes BVI people’s involvement only as consumers of these notebooks and limits discovery of the extent of notebook accessibility problems.

\begin{figure*}
	\centering
	\includegraphics[width=.75\linewidth]{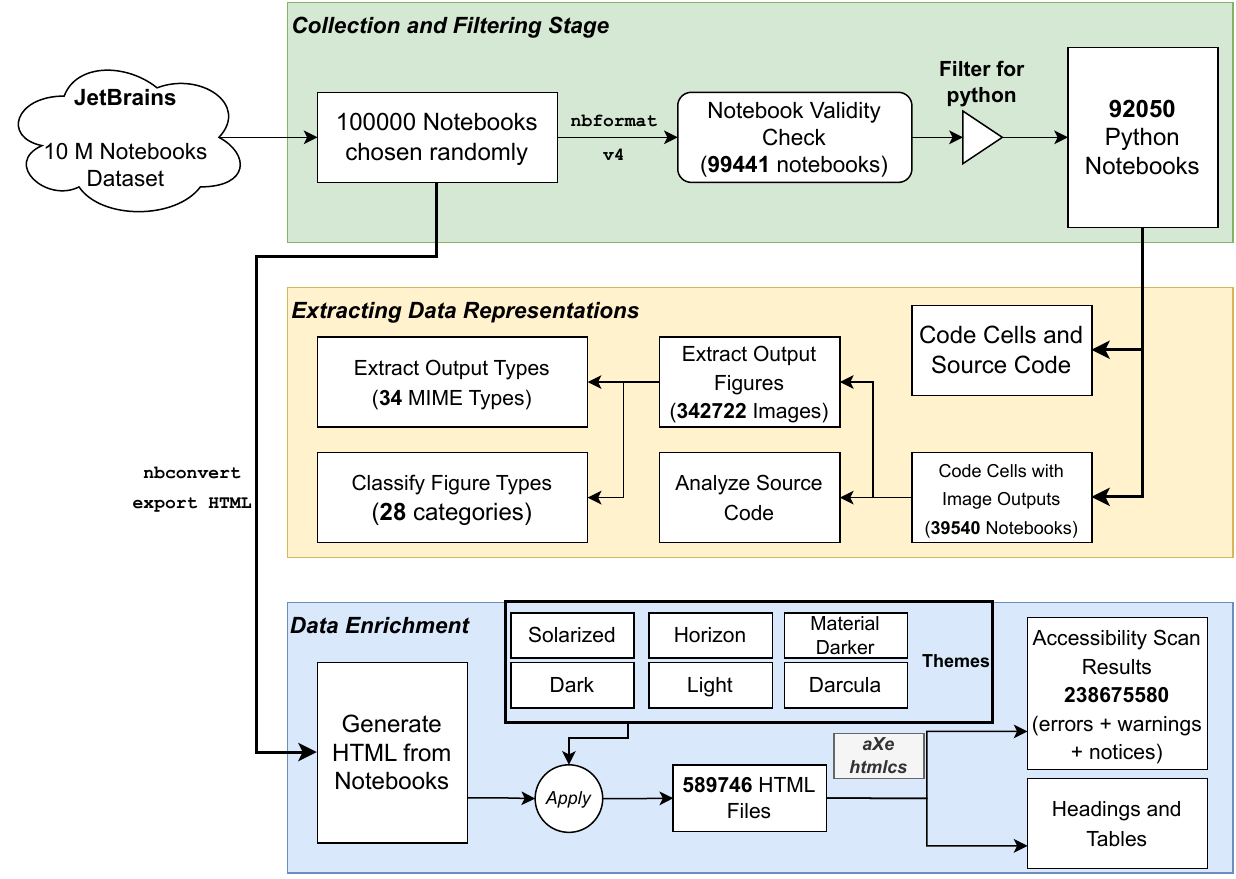}
        \Description[top-to-bottom Flowchart Diagram indicating the Data Processing Pipeline.]{The flowchart presents the three stages of the pipeline indicated by three large vertically aligned boxes color coded by the stage of the pipeline. Stage 1 is denoted by light green, stage 2 is denoted by pale yellow, and stage 3 is denoted by light blue. The first stage of the pipeline starts with choosing 100K notebooks randomly from the JetBrains dataset over which nbformat is run to check its validity. 99441 notebooks which pass the validity check are filtered for python resulting in 92050 python notebooks. These notebooks proceed to the next stage of the pipeline indicated by a yellow box. The source code from the code cells of the notebooks which programmatically generate images are selected. The resulting 39540 notebooks extract 342722 images which result in 34 output types, and classification into 28 categories at later stages of the pipeline. The source code from these notebooks are also further analyzed. The final data enrichment stage of the pipeline uses the 100K notebooks from the initial stage and uses nbconvert to export HTML from the notebooks. By applying six themes, we obtain 589746 HTML files over which the aXe and HTMLCS accessibility engines are run resulting in a total 238675580 errors, warnings, and notices. The same HTML files are used to extract heading and table related information.}
	\caption{Flowchart Diagram indicating the Data Processing Pipeline presented in \Cref{subsec:data_processing_pipeline},~\ref{subsec:extracting_data_representations}, and~\ref{subsec:data_enrichment}}
	\label{fig:data-processing-pipeline}
\end{figure*}

Below, we present a high level diagram of our data processing pipeline in~\Cref{fig:data-processing-pipeline}. We will describe the steps involved in our pipeline and provide details of the data along the way. We build our data processing pipeline to
\begin{inparaenum}
	\item collect and filter a randomized subset of the notebooks from the larger JetBrains dataset (\S\ref{subsec:data_processing_pipeline}),
	\item extract the required data representations from the filtered notebooks (\S\ref{subsec:extracting_data_representations}), and
	\item enrich the notebook data through analysis of transformed representations, used in the notebook distribution process (\S\ref{subsec:data_enrichment}).
\end{inparaenum}

\subsubsection{Notebook Validity Check}

The first step of the pipeline involves coercing the computational notebook files to the latest v4 specification  of the  Jupyter Notebook format using the \texttt{nbformat} tool to ensure validity~\cite{nbformat}. A notebook obtained in the dataset is considered as valid if the file has correctly formatted JSON content according to the specified Jupyter notebook format. This conversion resulted in a total of \validFileCountAfterNBConvert notebooks filtering out \filteredFileCountDuringNBConvert notebooks in the process due to conversion failures.  

\subsubsection{Filtering for Python Notebooks}
\label{subsubsection:processing_raw_ipynb_notebooks}

Building on previous works on notebook analysis that have established python as the most popular language used in computational notebooks~\cite{guzharina2020jetbrains, quaranta2021kgtorrent}, we filter the validated dataset obtained from the first step for notebooks written in python. By removing notebooks which are not written in python, we obtain 94722 notebooks, of which we further removed \numberOfNotebooksWithNoLanguageInfo which do not contain the language information in the metadata, resulting in a total of \numberOfPythonNotebooks python based notebooks.

\subsection{Extracting Required Data Representations}
\label{subsec:extracting_data_representations}

Computational notebooks store source code in `code cells' and the outputs from the execution of the code as its children formatted as `output cells'. Additionally notebooks also support the usage of markdown to display text which is formatted as a `markdown cell'. These cells cannot contain child attributes related to outputs as per the notebook format specification. We process each notebook and extract information about \begin{inparaenum}
	\item source code, and
	\item outputs.
\end{inparaenum}

\subsubsection{Code Cells from Notebooks}

We run the next stage of our data processing pipeline to extract information about the source code present within `code cells' in the \numberOfPythonNotebooks python notebooks. We extract the source code, and markdown text in this process in addition to computing the number of code and markdown lines and cells in a notebook.  Our analysis identifies \numberOfNotebooksWithSourceOutputs\ notebooks where at least one source code cell generates a graphical output into its corresponding output cell. We removed \numberOfNotebooksWithCodeOnly notebooks that only contained source code, and no accompanying outputs for the code segments from our analysis, leaving \numberOfNotebooksWithSourceOutputs\ notebooks for the next stage of the pipeline. 

\subsubsection{Output Figures}
\label{subsubsection:extracting_output_figures}

The Jupyter notebook format stores figures generated as a part of the code output as \texttt{base64} encoded strings with the metadata information of the corresponding Multipurpose Internet Mail Extensions (MIME Types) to identify the type of media output. By default, the notebooks support five image media mime types indicated by \texttt{image/bmp}, \texttt{image/gif}, \texttt{image/jpeg}, \texttt{image/png}, and \texttt{image/svg+xml}. Other media output types such as PDF generated from code are immediately converted to display as PNG format in a notebook with no additional extensions installed. We parse through all the \numberOfNotebooksWithSourceOutputs\ notebooks since these are the python notebooks containing at least one programmatically generated graphic in an output cell corresponding to a code cell. We convert the encoded base64 strings into image files and store them for further analysis resulting in \totalNumberOfImages\ total images.

\begin{figure*}
	\centering
	\begin{subfigure}[b]{.32\textwidth}
		\includegraphics[width=\linewidth]{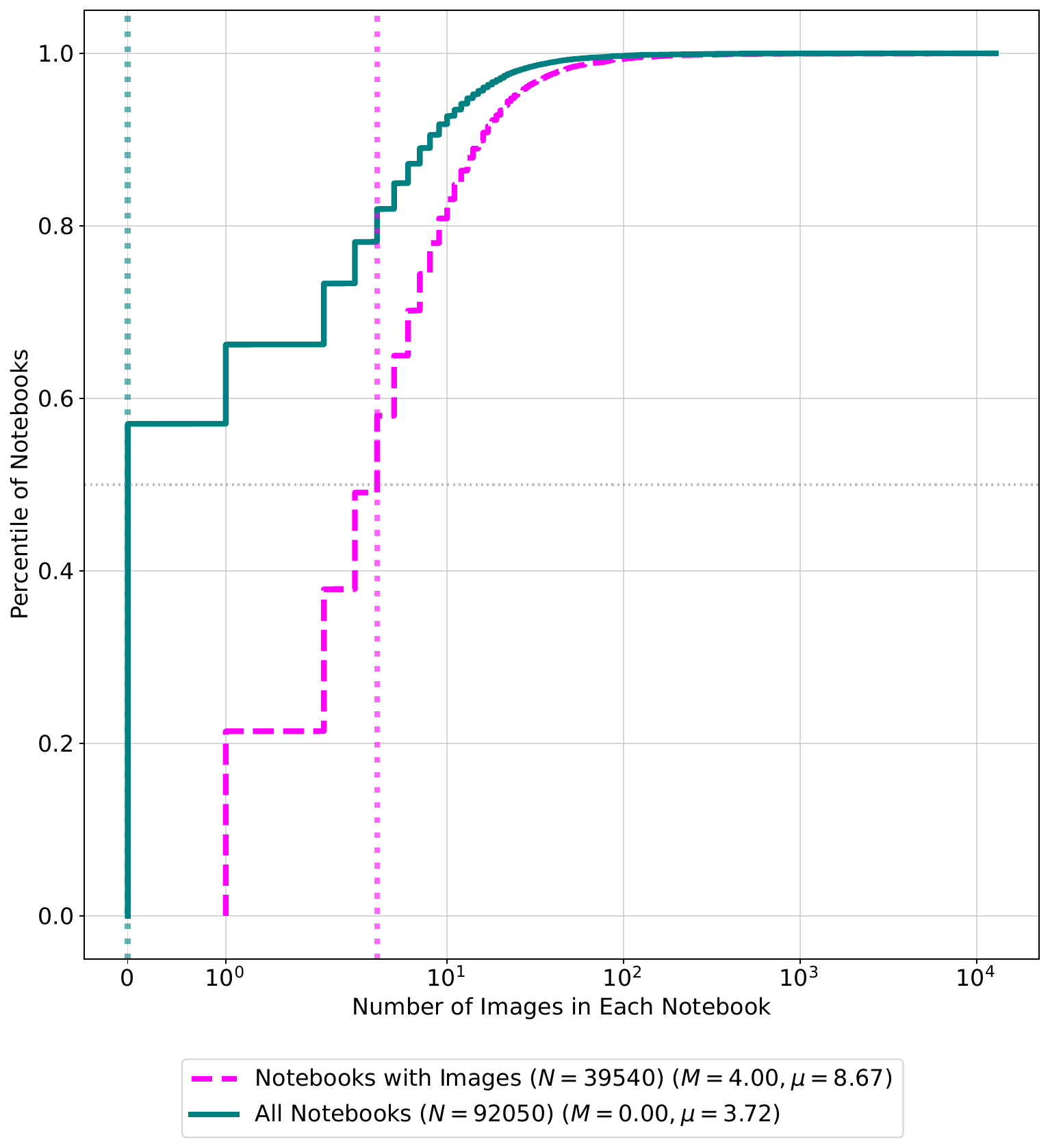}
            \Description[Comparative line charts showing Distribution of number of plots in each notebook]{Figure shows two lines indicating the distribution of the number of plots in each notebook. The median number of plots in 39540 notebooks containing images as 4.0 with mean of 8.67 images, and all 92050 python notebooks as 0 with a mean of 3.72 images.}
		\caption{CDF of number of plots in each notebook.}
		\label{fig:numPlots}
	\end{subfigure}
	\hfill
	\begin{subfigure}[b]{.32\textwidth}
		\includegraphics[width=\linewidth]{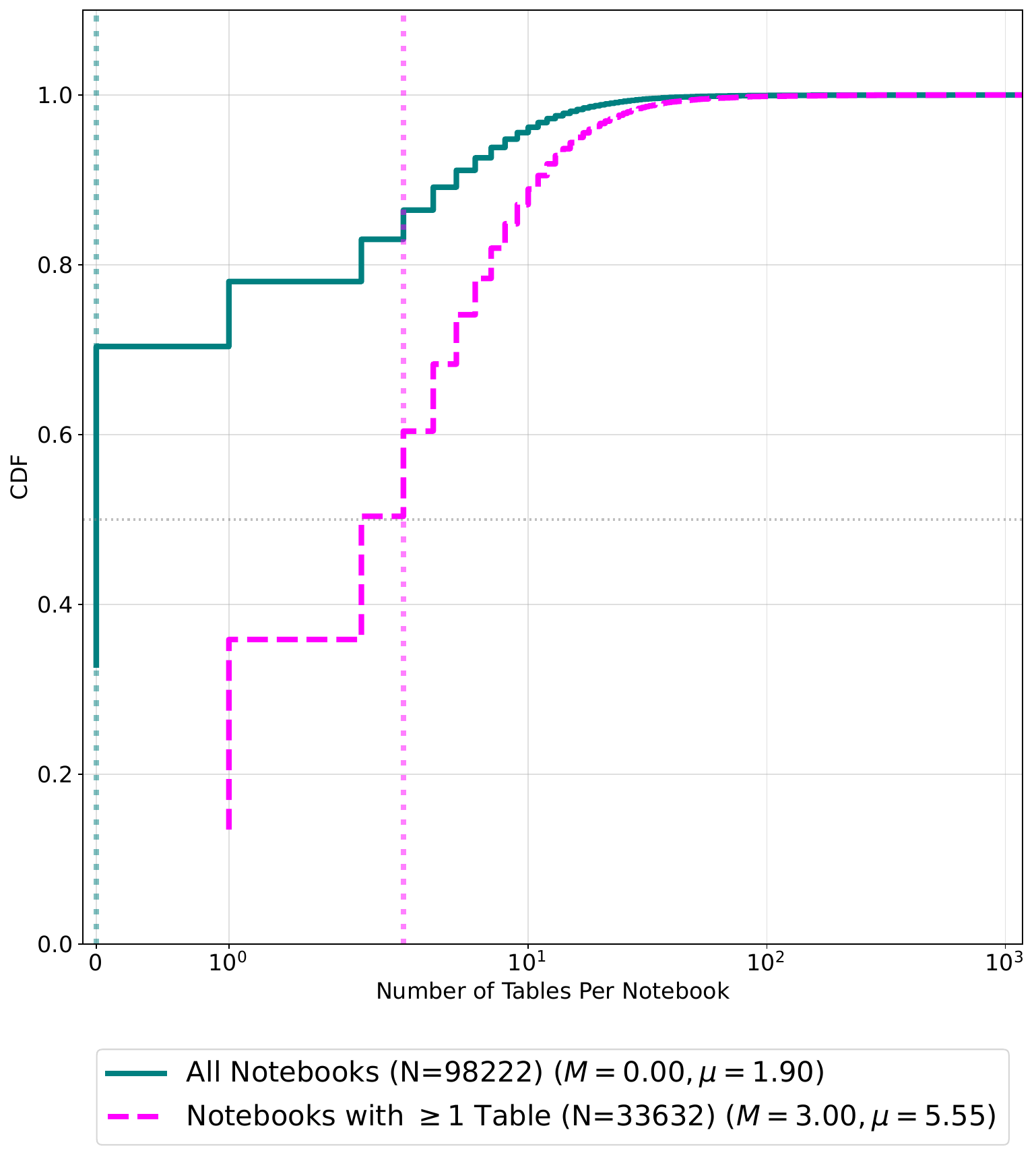}
            \Description[Distribution of Number of Tables in Notebooks]{Figure shows the distribution of the number of tables in notebooks. The solid teal line indicates that a notebook at the median contains no tables, with the average notebook containing 1.9 tables. However, the magenta line displaced to the right of the teal line in the figure indicates a median of 3 and a mean of 5.55 number of tables in the dataset.}
		\caption{CDF of number of tables in notebooks.}
		\label{fig:tablesInNotebooks}
	\end{subfigure}
\hfill
	 \begin{subfigure}[b]{.32\textwidth}
		\includegraphics[width=\linewidth]{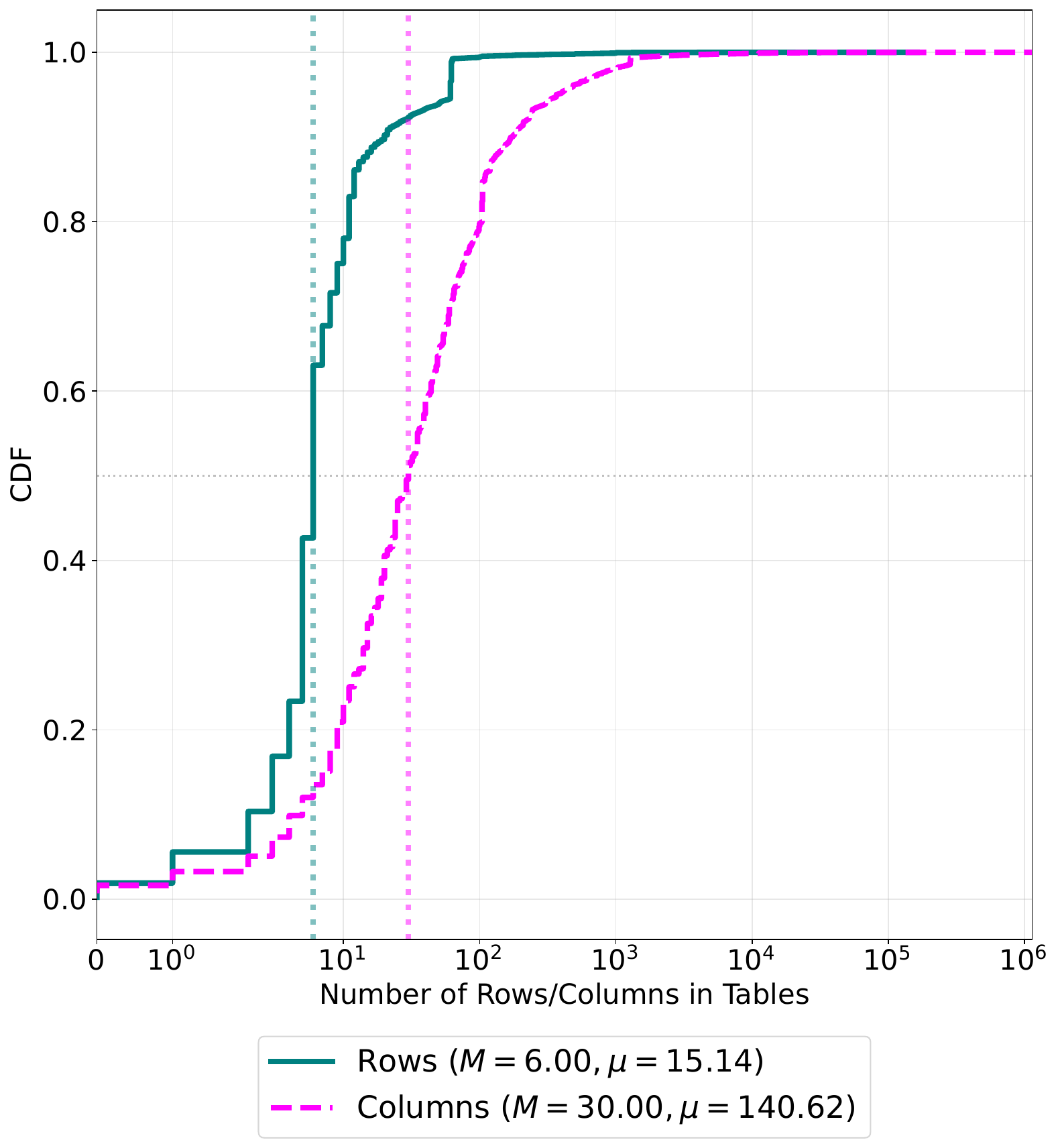}
            \Description[Distribution of Number of Rows and Columns in Tables]{Figure presents the distribution of the number of rows in tables using the solid teal line and the distribution of number of columns in the tables using the dashed magenta line. The median number of rows for a table is 6, with the average at 15.14, and the median number of columns for the tables are 30, with the average at 140.62.}
		\caption{CDF of number of table rows and columns.}
		\label{fig:numberOfRowsColumnsInTables}
	\end{subfigure}
        \Description[Notebook Characteristics in the Analysis]{The overall figure presents the characteristics of the notebooks in the analysis. The plots show cumulative distribution functions and use two colors teal and magenta with the solid lines colored in teal, and the dashed lines colored in magenta. All three figures use the same consistent color scheme and contain legends below the x axis.}
	\caption{Characteristics of the Notebooks in the Analysis. Plots show a cumulative distribution function (CDFs). The teal line in the CDF indicates the relationship between the  \numberOfPythonNotebooks valid python notebooks in our study and the number of plots they contain. The magenta line indicates the  distribution of \numberOfNotebooksWithSourceOutputs\ notebooks  which contain at least one image. The step nature of the curve compared to the continuous distribution is because of the discrete number of plots which could be included in a notebook. The vertical length of the line at each $x$ value indicates the percent of those notebooks in the dataset.
}
	\label{fig:characteristics}
\end{figure*}

As summarized in \Cref{fig:numPlots}, about \percentageOfNotebooksWithSourceOutputs notebooks have at least 1 programmatically generated figure. For notebooks that have such images, the median number of images in them  is \medianImagesPerNotebookFiguresOnly. 10\% of the notebooks contain over \pNinetyNumberOfImages images per notebook, with a long tail where 1\% of the notebooks contain over \pNinetyNineNumberOfImages images. The notebook with the most images contains \maxNumberOfFiguresInNotebook images. 

\subsubsection{Analysis of Code Syntax}
\label{subsubsection:extracting_code_markdown_cells}

The next step in our data processing pipeline  targets  the contents of the source code in the \numberOfNotebooksWithSourceOutputs python notebooks that contain at least one programatically generated graphical output. We chose Python because it is the most popular language used in notebooks and is studied in other large scale notebook analyses~\cite{choetkiertikul:2023:MiningNotebookCharacteristics,rule2018exploration}. This allows us to perform language specific code analysis by delving deep into the library and module ecosystem of the programming language, to understand accessibility gaps in popular tools, and make actionable recommendations to improve the accessibility of these notebooks.

We use the abstract syntax tree (\texttt{ast}) module in python to parse the contents of the source code in these \numberOfNotebooksWithSourceOutputs\ notebooks. We extract information about modules and functions being imported by notebook authors, and the functions being invoked within various cells of the notebook. This gives us information about the different libraries, and function calls frequently used by notebook authors.

Constructing the abstract syntax trees however is not as trivial as combining the code cells in a notebooks before running a parser, and  requires additional processing of the code in the notebooks. Notebook systems support Jupyter \textit{magics} --  a functionality provided by the kernel, that allow developers to call functions that would simplify some Jupyter operations. For example, the IPython kernel used by Jupyter, allows developers to use the \texttt{\%\% latex} command to render a cell in LaTeX and the \texttt{\%\% bash script magic} command to run a cell in bash as a subprocess. While these are valid operations in a Jupyter environment, they are not a valid part of the syntax of the python programming language and result in errors when parsing the syntax to construct an abstract syntax tree, even if the rest of the code in the cell is valid syntax.
Therefore, constructing abstract syntax trees requires the source code snippets to be processed to remove any magic lines. We removed source code lines which begin with the \texttt{\%} (percentage) special character, in addition to lines starting with \texttt{!} indicating the execution of a shell command, or ending with the help \texttt{?} operator. We run our parsers to construct the AST over the resulting code and extract information about the functions, and modules imported in the notebooks.

\subsubsection{Output Types}

The extensibility of Jupyter notebooks allows notebook authors to customize the story telling and consumption experience of the notebooks.
To understand how these customizations impact accessibility, we extracted three high level categories (\textit{application}, \textit{image}, and \textit{text}) based on MIME types used in notebooks. The top 5 output types presented in \Cref{table:top5-output-types} account for \topFiveOutputsCumulativePercentage of the outputs in the notebooks.  Notebooks included \numberOfApplicationMimeTypes different application types; \numberOfTypesOfTextTypes text output types (eg. HTML, \LaTeX, markdown, etc..,), and \numberOfTypesOfImageTypes image output types. Portable network graphics (PNG) images are the most commonly used image output formats making up 21.45\% of the outputs in the notebooks. The detailed and complete list of output types found in our dataset is presented in \Cref{appendix:output-types-list} in \Cref{appendix:types_of_outputs_in_notebooks}.

\begin{table*}[t]
    \caption{Top 5 Output Types in Notebook Account for \topFiveOutputsCumulativePercentage of outputs}
	%\begin{adjustbox}{width=\linewidth,center}
	\begin{tabular}{|lll|l|ll|}
		\hline
			\multicolumn{1}{|l|}{\textbf{Rank}} &
			\multicolumn{1}{l|}{\textbf{Category}} &
			\textbf{Output Type} &
			\textbf{Total} &
			\multicolumn{1}{l|}{\textbf{Percent}} &
			\textbf{Cumulative} \\ \hline
		\multicolumn{1}{|l|}{1} & 
			\multicolumn{1}{l|}{Text}        & 
			Plain               & 
			977328           & 
			\multicolumn{1}{l|}{61.72} & 
			61.72 \\ \hline
		\multicolumn{1}{|l|}{2} & 
			\multicolumn{1}{l|}{Image}       & 
			PNG                 & 
			339589           & 
			\multicolumn{1}{l|}{21.45} 
			& 83.17 \\ \hline
		\multicolumn{1}{|l|}{3} & 
			\multicolumn{1}{l|}{Text}        & 
			HTML                & 
			208170           & 
			\multicolumn{1}{l|}{13.15} & 
			96.32 \\ \hline
		\multicolumn{1}{|l|}{4} & 
			\multicolumn{1}{l|}{Application} & 
			Javascript          & 
			22842            & 
			\multicolumn{1}{l|}{1.44}  & 
			97.76 \\ \hline
		\multicolumn{1}{|l|}{5} & 
			\multicolumn{1}{l|}{Application} & 
			Jupyter JSON Widget & 
			14532            & 
			\multicolumn{1}{l|}{0.91}  & 
			98.67 \\ \hline
		\multicolumn{3}{|l|}{\textbf{Total Number of Outputs}} & 
			\textbf{1583255} & 
			\multicolumn{2}{l|}{}              \\ \hline
	\end{tabular}
	%\end{adjustbox}
	\label{table:top5-output-types}
\end{table*}

\subsubsection{Figure Types}
\label{subsubsection:classifying_figure_types}

To understand what type of figures are created in the notebooks, we  classify the figures into 28 different image type categories including Line Chart, Histogram, Box plot, Confusion matrix, Scatter plot and others (the full list can be found in ~\ref{fig:plotTypes}).
We classify the \totalNumberOfImages\ images obtained from our figure extraction phase of the pipeline, using a Fully Connected - Convolutional Neural Network (FC-CNN) combined with a Fisher-Vector Convolutional Neural Networks (FV-CNN)~\cite{song2017fisher} pretrained on the DocFigure dataset~\cite{Jobin:2019:DocFigure}. We run this inference on an AWS \texttt{p2.16xlarge} VM running 16 NVIDIA K80 GPUs, 64 vCPUs, and 732 GB of memory.

\subsection{Data Enrichment}
\label{subsec:data_enrichment}

The user experience to consume and author Jupyter notebooks often takes place through web interfaces. Developers customize their development environment experiences through themes, and many IDEs in addition to their defaults, provide a wide variety of themes -- catering to developer preferences. Often these same themes 
are carried to published notebooks. 

\subsubsection{Generating HTML from notebooks}
\label{subsubsection:generating_html_exports}

Since different themes can vary in their accessibility, we selected 6 popular themes including the defaults provided by Jupyter for publishing HTML exports. Our selections include \textit{solarized} -- a theme (originally written for Vim and made available now by various IDEs)~\cite{solarized}, \textit{darcula} (for ZSH originally, and used extensively by JetBrains)~\cite{darcula}, \textit{horizon} (default by VSCode)~\cite{horizon}, the \textit{material darker} theme~\cite{material-darker}, and the default \textit{dark} and \textit{light} theme options supported by Jupyter~\cite{jupyter}. To assess the accessibility of notebooks in these web interfaces, we generate HTML exports of notebooks in our dataset with these six popular themes applied.

We use the open source \texttt{nbconvert} tool that supports converting notebooks into publishable HTML and other formats. \texttt{nbconvert} allows users to select themes and specify other parameters to control the generation of HTML output and is the de-facto tool integrated into computational notebooks providing various export format capabilities. 
Once the conversion is complete, we export the notebooks into standalone HTML files which we then serve through a web server. 
All \numberOfNotebooks notebooks were exported as HTML using \texttt{nbconvert}, producing \totalNumberOfHTMLFilesAcrossThemes HTML files, \numberOfHTMLFilesPerTheme per theme.

\subsubsection{Accessibility scans}
\label{subsubsection:accessibility_scans}

HTML user interfaces are typically evaluated using accessibility testing and evaluation engines which are software programs that evaluate the content, design of the interface, and check their ability to satisfy various established accessibility guidelines. We perform accessibility scans using aXe and HTML Code Sniffer (HTMLCS) together by deploying a self hosted version of the \textit{pa11y} accessibility scanning infrastructure configured to use both engines~\cite{pally}. \textit{pa11y} is an LGPL licensed open source tool that tests web pages by executing a chromium process and providing the ability to run multiple accessibility engines via the same tool. We modified \textit{pa11y}, which was last modified by the maintainers in October 2022, by building on the community's work to make the project compatible with the latest Node.js $\geq$ v18. Our changes expose the ability to run multiple accessibility engines in their webservice subproject~\footnote{\url{https://github.com/pa11y/pa11y-webservice/pull/144}} and additionally involved fixing various security vulnerabilities~\footnote{\url{https://github.com/pa11y/pa11y-webservice/pull/145}}, and overcoming engineering debt. 
We are currently working towards contributing these changes to pa11y.

For all \totalNumberOfHTMLFilesAcrossThemes HTML files across all themes, we extracted the type of violation (\textit{error}, \textit{warning} or \textit{notice}), the accessibility engine that detected it (Axe or HTMLCS), the specific code corresponding to the violation provided by the engines, and the selector (HTML node where the violation was detected). We enrich our dataset by attaching this information to the name of the notebook and theme being tested. We found a total of \totalNumberOfScanResults combined errors, warnings and notices across all notebooks.

\subsubsection{Web Semantics}

While accessibility scanners provide information about standardized accessibility errors, they do not provide more nuanced information such as the size of tables and the presence of headings that can be critical to understand the screen reader glanceability and navigability of notebooks. To gain this understanding, we use LXML, a library that enables efficient parsing of the HTML DOM tree, to process the HTML outputs of notebooks. We picked the light theme and used the  \numberOfHTMLFilesPerTheme HTML files corresponding to it for this processing. 
We extracted information about the type of cell, presence of images, alternative text (using the alt attribute of the \texttt{<img>} tags), information about tables (using the \texttt{<table>}, \texttt{th}, and \texttt{td} tags), presence of links (using the \texttt{<a>}), and various heading levels (using the  \texttt{<h1 - h6>} tags) along with the file name, and the location of the cell in the file.

Our observations indicate that \percentageNotebooksWithNoTables of the notebooks do not contain any tables in their outputs. Among the remaining, a notebook contains \medianNumberOfTablesInNotebooks (median) / \meanNumberOfTablesInNotebooks (mean); however the distribution has a long tail with the maximum at \maximumNumberOfTables tables and 1\% of the notebooks containing over \lastPercentileNumberOfTables tables.  Among the \percentageNotebooksWithTables of notebooks in our dataset which contain tables, we extract additional metadata to identify the structural shape of the tables. The median table rendered in the output cells of the notebooks contains \numberOfMedianRowsInTables (median) / \numberOfMeanRowsInTables (mean) rows, and \numberOfMedianColumnsInTables (median) / \numberOfMeanColumnsInTables (mean) columns. \Cref{fig:numberOfRowsColumnsInTables} indicates the distribution of the number of rows and columns for all tables identified in our dataset. 
The largest table in the notebooks in our dataset contains \largestTableContainsRows rows and \largestTableContainsColumns columns indicating a maximum of \largestTableMaxCellCount cell values.

\subsection{Manual Screen Reader Testing}
\label{ssec:ManualTesting}

To investigate a sample of accessibility issues identified by our automated testing, and to verify if notebooks were glanceable,  two research team members experienced with using a screen reader opened a selection of notebooks in   screen reader and browser combinations, and verified if a screen reader user will be able to get to all headings, images, and tables present in those notebooks. As we noticed the BVI researcher's screen reader crash several times during the exploratory phase of our research, we hypothesized size of the notebooks to play a role in causing these crashes. We identify a list of notebooks meeting different size criteria based on the percentiles of the file sizes in our dataset. While one researcher (also a BVI screen reader user) performed the tests and reported the counts in each notebook, the second researcher visually verified the reporting, noting the observations. We performed these tests with Microsoft Edge, Google Chrome, and Firefox Nightly with their accessibility cache improvements enabled~\cite{CachetheWorld2022Asa} with JAWS and NVDA 2023 on a Windows computer running Windows 10. We performed our VoiceOver tests on a Mac with VoiceOver using both Safari, and Chrome. We did not test our notebooks with Firefox on the Mac since the accessibility cache improvements for Firefox were not yet available for the Mac OS. Similarly, we did not test using Safari on Windows since Apple ended support for it in 2015.

\section{Results}
\label{sec:results}

Recall that the goal of our investigation is to understand the accessibility of data artifacts, authoring experiences, and infrastructure to work with computational notebooks to BVI users.
Given the overall concerns about visualization and data accessibility discussed in \cref{background:accessibility_of_data}, we must understand how accessible these are in notebooks. Thus, we begin our analysis by exploring how accessible the data artifacts in notebooks are to BVI users.
Given the importance of navigation, and information seeking/glanceability to  IDE accessibility, it is critical that we assess this in notebooks as well. Thus, our analysis  explores the proper use of headers and other landmarks that improve navigation and glanceability for screen reader users.
Uniquely, notebooks are not just a programming tool, they are also a communication platform that can be customized and themed  to generate final ``documents'' in various formats. If these documents are not accessible, the consumer experience will be impacted. Thus, our analysis investigates the impact of the tools used to customize and distribute notebooks on their accessibility.
To answer our research questions, we developed metrics  which represent optimistic \textit{upper bounds}, meaning that the presence of accessibility in notebooks is likely much smaller than our findings in this paper indicate.

\subsection{Accessibility of Data Artifacts}
\label{ssec:artifacts}

At the heart of data analysis is the data, and that is typically explored through a combination of text summaries, graphical representations, and tables. These latter two artifacts -- \textit{graphics} and \textit{tables}, that are essential to the data story telling process used by notebook authors, may have important implications for accessibility and help us answer the first research question \textit{RQ1} we set out to answer.  A truly accessible notebook should support advanced, accessible and dynamic visualization techniques~\cite{kim2021accessible}. However there are common, basic requirements for making static images and charts accessible~\cite{Elavsky:2022:Chartability}, which guided the development of our optimistic metrics for evaluating artifact accessibility. 

\begin{description}[style=unboxed,leftmargin=0cm,itemsep=1em]
\item[Presence of ALT text]: A meaningful \textit{ALT} text attribute should accompany visualizations. We measure the presence or absence of ALT text in programatically generated images and analysed the alt-texts found in images that may have been manually added by notebook authors. This is an  optimistic measure  because the presence of ALT text does not mean it is descriptive or helpful for accessibility -- which depends heavily on the quality of the text provided.
Only .19\% of images in our data have ALT text whose word frequency we measure and present detailed results of in \Cref{subsec:accessibility_of_images} but we do not measure ALT text quality.
\item[Figures followed by tables]: Without ALT text, a figure can still be somewhat accessible if it is followed by a table with the equivalent data visualized in the figure. This is optimistic because such tables may not be accessible, or may not be related to the figure. We present the detailed results in \Cref{subsec:comparitive_ordering}.
\end{description}

\subsubsection{ALT text for Static Images and Charts}
\label{subsec:accessibility_of_images}

\begin{figure*}[ht!]
    \centering
     \begin{subfigure}[b]{.35\linewidth}
        \includegraphics[width=\linewidth]{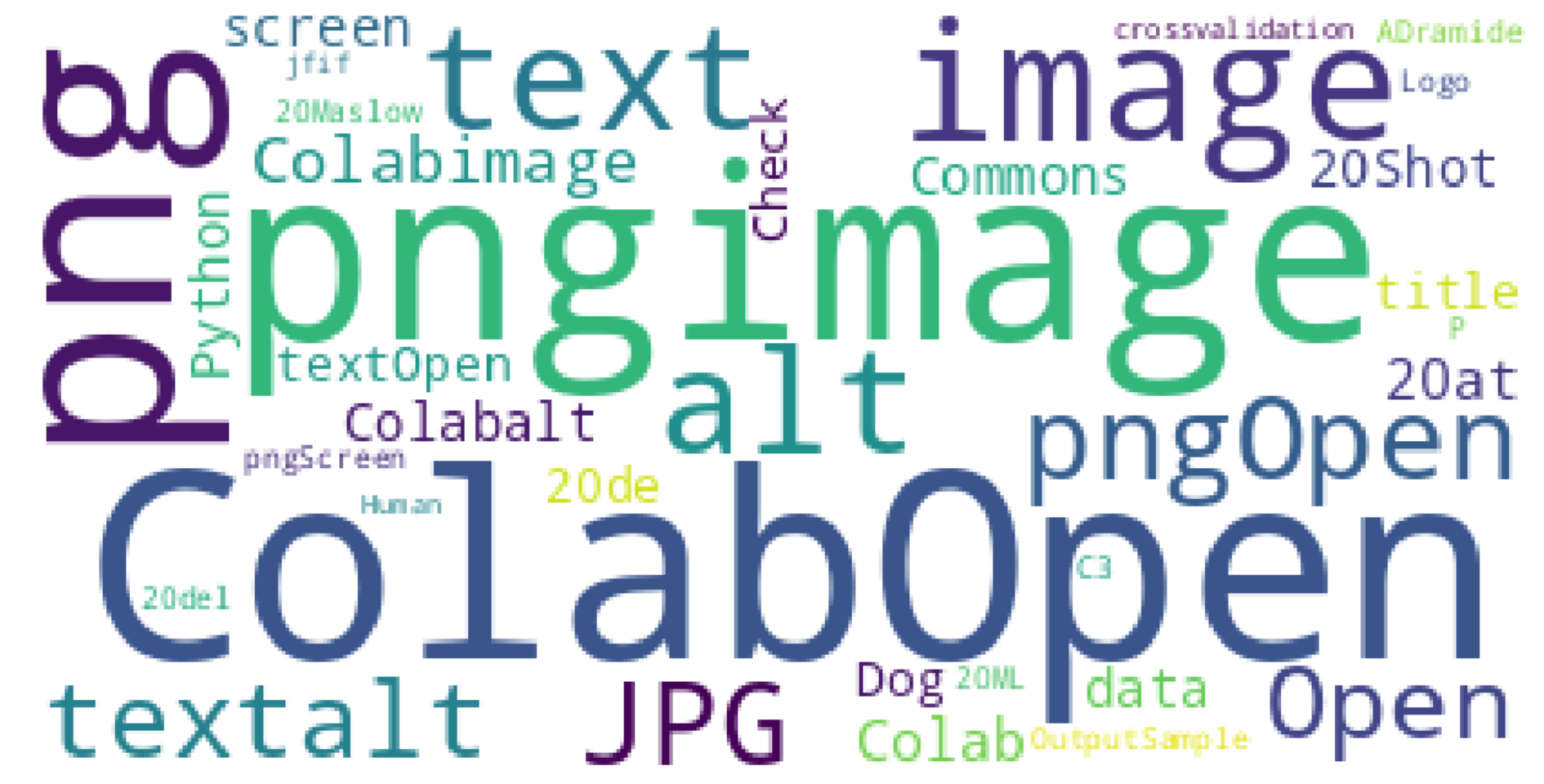}
        \Description[Wordcloud of ALT text in Images found in Notebooks]{Figure presents a word cloud of the alt text descriptions present in the notebooks which mostly contain words that are not helpful to screen reader users such as the word `image', `alt', or `png'. The words Colab and Open also occur at high frequencies dominating the word cloud's central display.}
        \caption{Wordcloud of Image ALT-Texts in Notebooks.}
        \label{fig:alt-text-word-cloud}
    \end{subfigure}
    \hfill
    \begin{subfigure}[b]{.60\linewidth}
        \includegraphics[width=\linewidth]{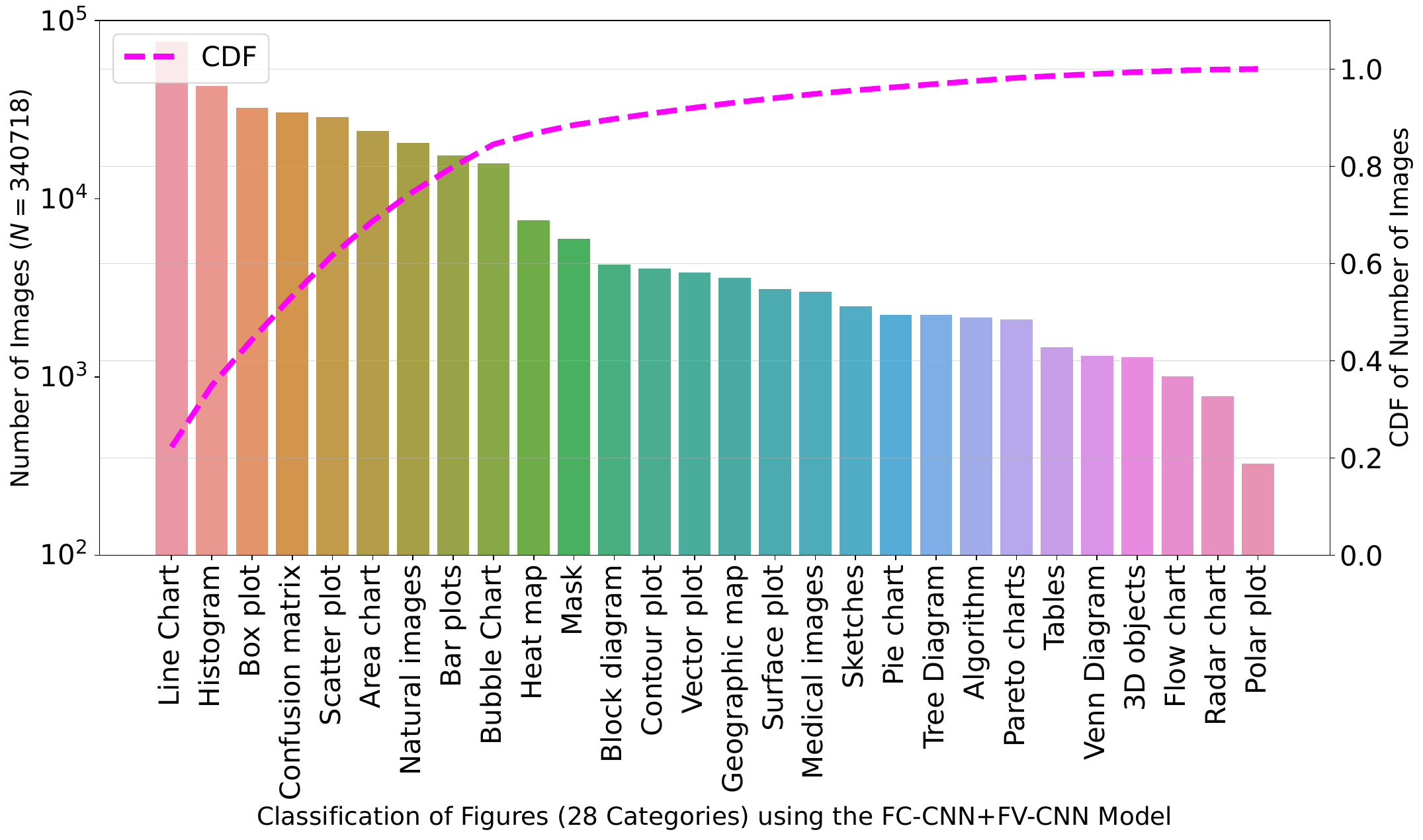}
        \Description[Classification of \totalNumberOfImages images used in \numberOfNotebooksWithSourceOutputs notebooks.]{Figure 3b is a histogram whose x axis contains the categories of classification of images in the dataset after inference through the FC-CNN+FV-CNN model, and y axis represents the total counts of images over a log scale. The sum of all vertical bars indicating different categories ranked in descending order of usage popularity result in the total 340718 images. The figure contains an alternate y axis to the right side of the graph indicating the CDF of the number of images over all categories. Over 75 percentage of the images are classified into 7 types (Line chart, histogram, box plot, confusion matrix, scatter plot, area chart, and natural images).}
        \caption{Classification of \totalNumberOfImages\ images used in \numberOfNotebooksWithSourceOutputs\ notebooks into 28 categories.}
        \label{fig:plotTypes}
    \end{subfigure}
    \Description[Types of visualizations used in notebooks]{ The figure contains two subfigures indicating the information extracted from the visualizations and images in the notebooks.}
    \caption{Types of visualizations used in notebooks and analysis of most frequent words used in alternate text descriptions.}
\end{figure*}

Static images and charts are present in \percentageOfNotebooksWithSourceOutputs (\numberOfNotebooksWithSourceOutputs)  of the  \numberOfPythonNotebooks python based notebooks, most of which are PNG files (\Cref{table:top5-output-types}). Notebooks with these artifacts contain a median of \medianImagesPerNotebookFiguresOnly figures as shown in \Cref{fig:numPlots}. We consider a figure to be programmatically generated if the code cell in the notebook contains a mapped output section indicating the result of the execution of the cell and containing an image type. The vast majority of the programmatically generated images (N=\totalNumberOfImagesWithoutAlt) do not have associated alternative text. Of the \totalNumberOfImagesWithAlt containing the alt text information in the notebooks, only \totalNumberOfImagesWithAltFromCode image is programmatically generated from code~\footnote{We investigate the one image with alt text attribute generated programmatically in the notebook but cannot reproduce the result \texttt{36f17428463ea6a6a75d9e064713a3039f381037.ipynb}}, while \totalNumberOfImagesWithAltFromMarkdown others are specified through the \texttt{description} attribute in markdown images included using the syntax `\texttt{![description](path\_to\_image)}'. 
In \Cref{fig:alt-text-word-cloud}, we present a word cloud of the alt descriptions we found indicating that they were mostly meaningless in their current use. The most dominant words `Open' and `Colab' come from the markdown included interaction button for opening notebooks with Google Colab which is included by some notebook authors when publishing their notebooks. The word cloud also indicates poor usage of alt text whenever used in markdown with most descriptions lackadaisically referring to the graphic included as `image', `png', or `alt'.

\paragraph{Types of charts and libraries used}
\label{subsubsection:understanding_types_of_charts}

\begin{figure}[!ht]
    \centering
    \includegraphics[width=\linewidth]{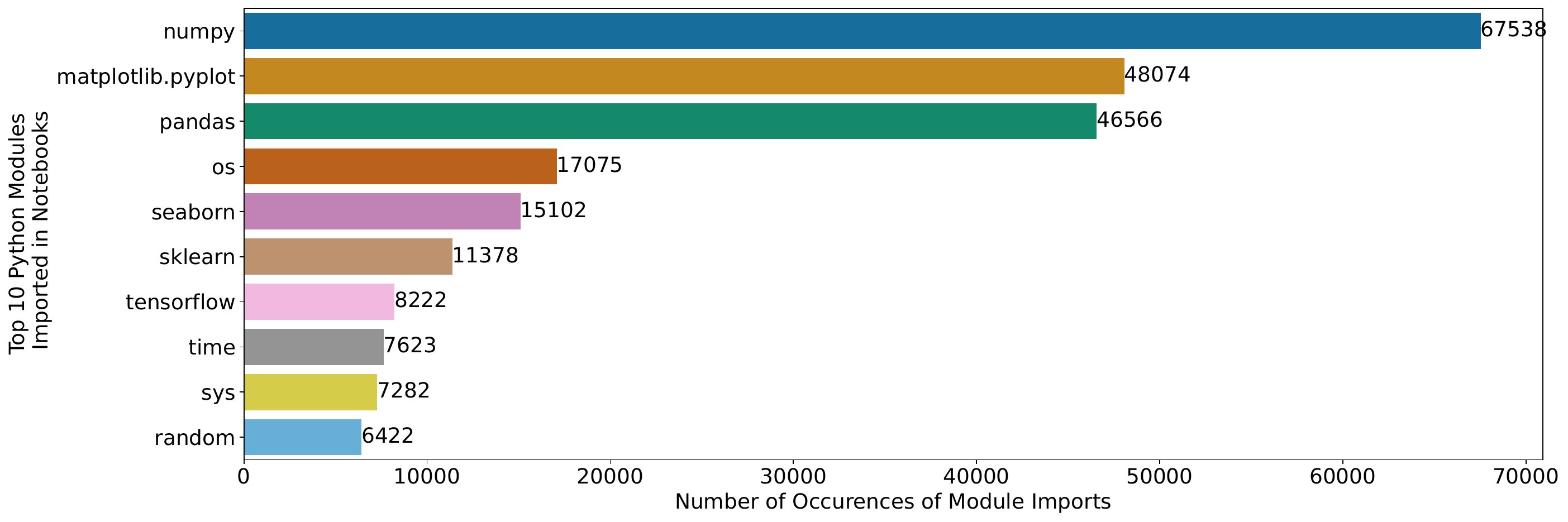}
    \Description[Top 10 popular python modules imported in notebooks ranked by their usage frequency]{The figure contains a horizontal bar plot with the x axis indicating the number of occurrences of module imports, and the y axis indicating the top 10 imported python modules. The horizontal bars use colors from the colorblind palette provided by seaborn and indicate numpy, matplotlib.pyplot, pandas, os, seaborn, sklearn, tensorflow, time, sys, and random with their respective counts.}
    \caption{Top 10 popular python modules imported in notebooks ranked by their usage frequency}
    \label{fig:importNums}
\end{figure}
\begin{table*}[!ht]
    \caption{Top 10 Most Invoked Function calls in Python Notebooks with Figure Outputs (excluding built in functions)}
	\begin{tabular}{|l|l|l|l|l|l|}
		\hline
		\textbf{Rank} &
		\textbf{\begin{tabular}[c]{@{}l@{}}Rank\\ (incl. built ins)\end{tabular}} &
		\textbf{Function Targets} &
		\textbf{Module} &
		\textbf{Function Call} &
		\textbf{Count} \\ \hline
		1  & 4  & Visualization & \texttt{matplotlib} & \texttt{plt.plot}     & 70575 \\ \hline
		2  & 5  & Visualization & \texttt{matplotlib} & \texttt{plt.show}     & 70454 \\ \hline
		3  & 6  & Visualization & \texttt{matplotlib} & \texttt{plt.title}    & 49998 \\ \hline
		4  & 8  & Data          & \texttt{numpy}      & \texttt{np.array}     & 46360 \\ \hline
		5  & 9  & Visualization & \texttt{matplotlib} & \texttt{plt.figure}   & 45797 \\ \hline
		6  & 10 & Visualization & \texttt{matplotlib} & \texttt{plt.xlabel}   & 42430 \\ \hline
		7  & 11 & Visualization & \texttt{matplotlib} & \texttt{plt.ylabel}   & 41693 \\ \hline
		8  & 12 & Data          & \texttt{pandas}     & \texttt{pd.DataFrame} & 32144 \\ \hline
		9  & 13 & Data          & \texttt{pandas}     & \texttt{pd.read\_csv}  & 30858 \\ \hline
		10 & 14 & Visualization & \texttt{matplotlib} & \texttt{plt.legend}   & 23228 \\ \hline
	\end{tabular}
	\label{table:top-10-function-calls}
\end{table*}

If we understand what types of charts and figures  are most used by notebook authors, this can help guide and prioritize research in accessible visualization for data analysis. 
The usage of the pre-trained FC-CNN+FV-CNN classification model (\Cref{subsubsection:classifying_figure_types}) over the programmatically generated images found in our data (N=\totalNumberOfImages) reveals line charts as the most common type of programmatically generated visualization (22.26\%), followed by histograms (12.61\%), and boxplots (9.45\%). Figure~\ref{fig:plotTypes} contains a histogram of the most popular types of figures found in the analyzed notebooks ranked from most widely used to the least widely used. Seven types of charts make up roughly 75\% of the figures in the dataset: Line chart, histogram, box plot, confusion matrix, scatter plot, area chart, and natural images. Some of these are understudied in the accessible visualization literature, which primarily focuses on bar, line, \& pie charts, and scatter plots~\cite{kim2021accessible}. Histograms, box plots, confusion matrices, and area charts represent important areas to make more accessible in  future research.

We also analyzed the python \texttt{import} statements and function call invocations used in the \numberOfNotebooksWithSourceOutputs\ notebooks with images, to identify the most popular charting library used by notebook authors (\texttt{matplotlib}, followed by \texttt{seaborn} as shown in \Cref{fig:importNums}). The seaborn library is an enhanced wrapper over the underlying matplotlib libraries. Additionally, through code analysis and tracing the function calls we identified the most popular functions used by notebook authors and present them in \Cref{table:top-10-function-calls}. 
Seven of these 10 popular functions produce visualizations through the matplotlib module, while the others target data processing through modules like \texttt{numpy} and \texttt{pandas}. Given this, we investigated matplotlib's capabilities to understand support for alt text. However, despite its importance in the python community and extensive use in computational notebooks, scientific publishing, and other media, Matplotlib does not support  embedding alt text descriptions for the generated graphics. Even with static figures, it is possible to include ALT text -- PNG, the most popular image format standard used by authors (\Cref{table:top5-output-types}), supports embedding image descriptions in metadata.

\subsubsection{Comparative Ordering of Tables and Figures}
\label{subsec:comparitive_ordering}
Unless a chart's summary, context, and alt text convey all relevant information contained in the chart, which our investigation of ALT text shows is clearly lacking in this data (\Cref{subsec:accessibility_of_images}), a data table or data summary is a necessary accompaniment to that chart~\cite{Elavsky:2022:Chartability}.  
With the use of data processing libraries like \texttt{pandas}, it is relatively straightforward to also render a slice of the dataset represented in a chart as an HTML formatted table, or to provide a description in the surrounding code or markdown cells. To estimate the prevalence of these accessibility best practice guidelines, we look for code cells that programmatically output an image and filter those which
\begin{inparaenum}
    \item do not contain a heading immediately after the current output cell -- since they typically indicate a change in the context~\cite{choetkiertikul:2023:MiningNotebookCharacteristics},
    \item have a markdown cell before or after the current cell potentially indicating necessary context, or 
    \item have a cell with a table (either programmatically generated or via markdown) immediately before or after the cell with the image.
\end{inparaenum}

Of the \totalCellsWithOutputGraphicsFromCodeCells code cells across all notebooks that contain images, we find that \totalCellsWithOutputGraphicsAndNoHeadingAfter (76.44\%) cells meet our criteria for possibly being related to the image. Of these, \totalCellsWithOutputGraphicsAndNoHeadingAfterContainingMarkdownBeforeOrAfter (37.13\%) cells contain markdown text (indicating possible explanations of the image) and \totalCellsWithOutputGraphicsAndNoHeadingAfterContainingTablesBeforeOrAfter (8.18\%) cells contain a table.

The \totalCellsWithOutputGraphicsAndNoHeadingAfterContainingTablesBeforeOrAfter cells surrounded by a data table in the neighboring cells are found in \numberOfNotebooksWithPotentialRelatedTables \textit{notebooks} in our dataset, while the \totalCellsWithOutputGraphicsAndNoHeadingAfterContainingMarkdownBeforeOrAfter cells containing markdown text after an image are present in 19028 notebooks. The existence of markdown cells in the neighboring cells but no heading in the cell immediately after however does not immediately imply relevance to the figure generated in the current code cell since they could also contain markdown included images. The existence of both markdown content, and supporting tables indicate the most accessible representations of the image -- containing both the relevant tables and a description.

Only \totalCellsWithOutputGraphicsAndNoHeadingAfterContainingMarkdownAndContainsTablesBeforeOrAfter (1.23\%) cells in \numberOfNotebooksWithPossibleManualTablesAndSupportingExplanation (4.53\%) notebooks with figure outputs meet this criteria and contain both markdown and tables in their neighboring cells making those cells relatively more accessible when navigating and attempting to understand the notebook.

Simply including tables after charts is a low bar. It is also important that those tables are usable with a screen reader. 
Despite the inclusion of tables in the notebooks, too many rows and columns can affect screen reader navigation, resulting in users losing context or requiring too many key strokes to skip the tables or interact with elements in the table. While there is no defined threshold for the maximum number of rows or columns, screen reader users typically lose context when navigating large tables~\cite{Wang:2022:webDataTables}. The median row count (6 rows) is identical to the number of default rows printed by \texttt{pandas.head()} (one header row, and 5 data rows). The number of columns however grows rapidly as shown in \Cref{fig:numberOfRowsColumnsInTables}. For example, the largest table in our dataset contains \largestTableContainsColumns columns and \largestTableContainsRows rows and may be impossible to glance with a screen reader. An average table present in the notebooks in our study contains 140 columns and 15 rows resulting in over 2100 cells indicating the very high number of keyboard interactions needed by BVI users to understand and navigate the tables.

In summary, we find that images and tables -- data artifacts found in notebooks --  may not contain the necessary information for them to be accessible to BVI notebook users. The presence of tables and text descriptions indicate the possibility for these data artifacts to be accessible and open up avenues for future investigations. However, notebook authors following this accessible practice need to also provide descriptions and use tables with  sizes considering  screen reader accessibility.

\subsection{Navigability and Accessibility for Authors}
\label{ssec:glanceability}

An understanding of the practices of notebook authors  can guide our approach to making notebooks more accessible. Current understanding of code glanceability challenges for BVI people (\eg~\cite{Potluri:2018:CT}) does not account for code representations which are interwoven with rich representations of data and web semantic structures -- such as headings and tables supported by computational notebooks. Since Jupyter notebooks support markdown,  a markup that is rendered using web semantics,  authors have a great deal of control over how structural information is conveyed in the notebooks. For example, notebook authors control presence of semantic elements such as headings, tables, links, and figures. Screen readers  typically support single key navigation among headings of different levels, links, tables, and many others~\cite{nvaccess}.  This can allow screen reader users to skim through a notebook, and quickly understand its structure.   
To explore notebook glanceability through web semantics helping answer the \textit{RQ2} we set out to explore, we define the following optimistic estimates of accessibility:
\begin{description}[style=unboxed,leftmargin=0cm,itemsep=1em]
    \item[Navigability of Notebooks:] One simple but important aspect of proper heading use to use a level 1 heading (H1) in the first cell of every notebook, enabling screen reader users to easily find the start of  notebook content~\cite{notebook-authoring-checklist-iotaschool}. This is an optimistic estimate because the heading should contain text that is relevant to highlighting the topic of the notebook, and  it is only one of several rules that should be followed to properly support navigability with headers. We present detailed findings in \Cref{sssec:headings}.
    \item[Finding landmarks:] \textit{Tables} are an important and accessible way of representing data to a screen reader user, especially given the lack of ALT text we found in visualizations. They also make it easy to quickly gain some insights about a notebook, because screen reader users can jump through tables using the single-key navigation similar to headings.
    Thus, a second simple and optimistic metric is whether a notebook has a table in it. Only  \percentageNotebooksWithTables notebooks have data tables and we present detailed results in \Cref{sssec:tables}. 
\end{description}

\subsubsection{Navigability of Notebooks}
\label{sssec:headings}

\begin{figure*}[ht!]
    \centering
    \begin{subfigure}[b]{.48\linewidth}
    	\includegraphics[width=\linewidth]{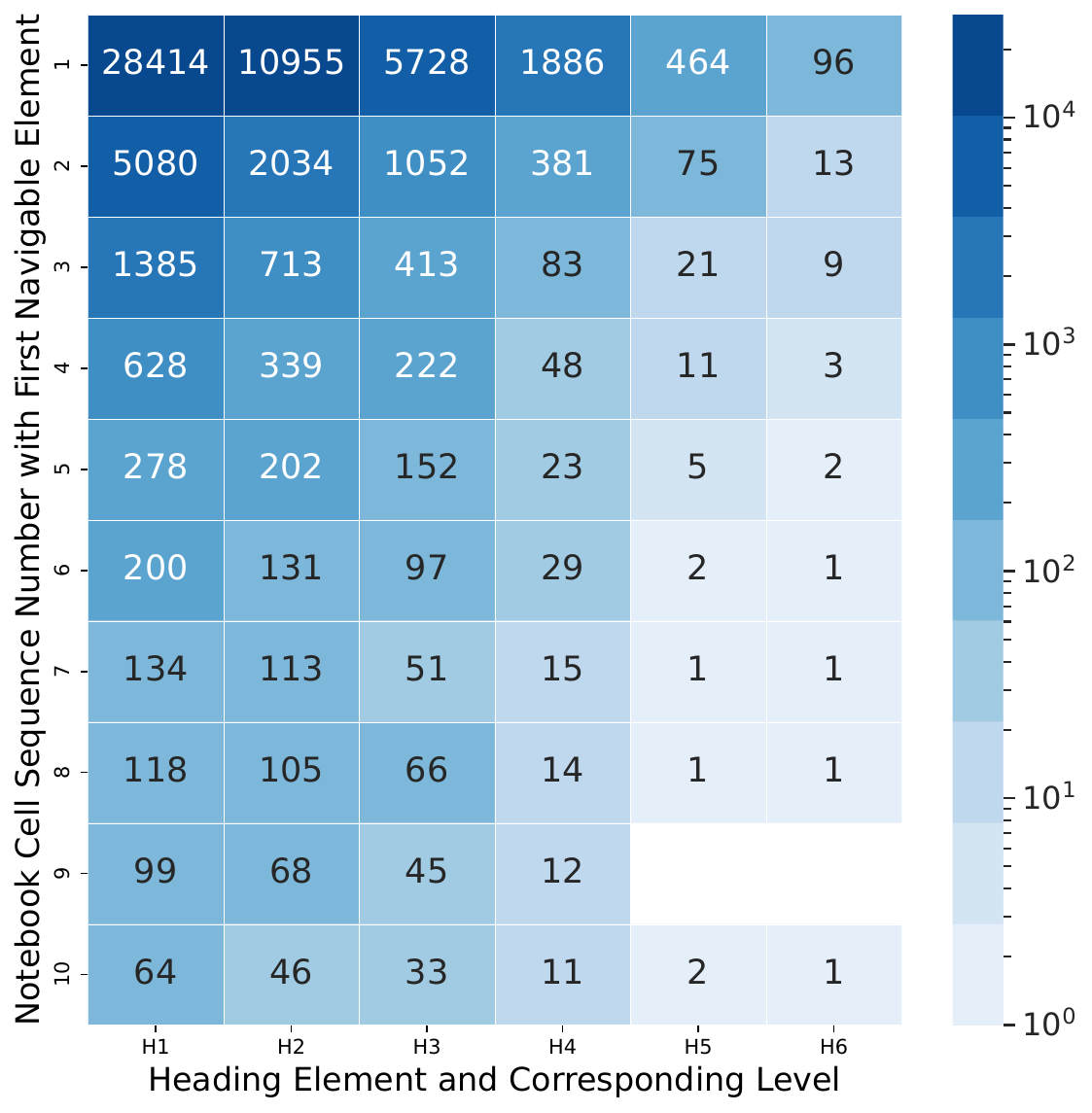}
            \Description[First Navigable Heading Element]{Figure represents a heatmap containing the frequency of headings existing at a cell location of the first occurrence of a heading and its corresponding level. The figure is color coded with the blues  color map where the x axis indicates the 6 heading levels and the y axis indicates the numbers starting from 1 to 10 representing the first 10 cells of a notebook. The heatmap contains the highest number 28414 coded as dark blue at the top left corner and is surrounded by smaller numbers in lighter gradients of blue to its right and bottom indicating the occurrence of a heading level 2 at cell 1 being more than the occurrence of heading level 1 at cell 1. There are two holes indicating no data for H5 and H6 at cell 9.}
    	\caption{First Navigable Heading Element}
    	\label{fig:hlevel_heatmap}
    \end{subfigure}
    \hfill
    \begin{subfigure}[b]{.48\linewidth}
    	\includegraphics[width=\linewidth]{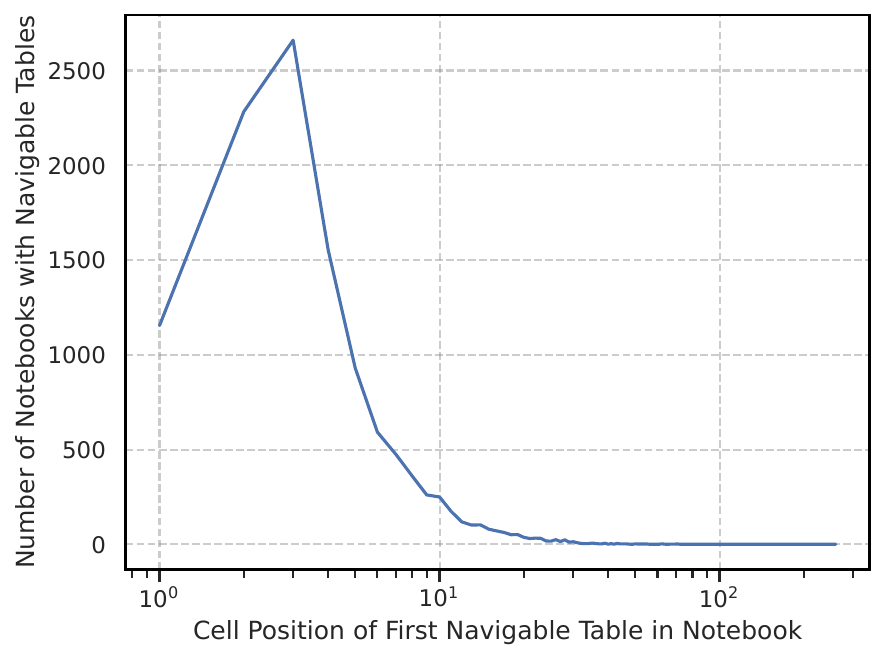}
            \Description[Position of First Navigable Table]{Figure represents a line chart indicating the frequency of the position of the first navigable table in a notebook. The line chart's $x$ axis is a log scale representation of the cell position in a notebook, and the $y$ axis indicates the number of notebooks with navigable table elements. The line starts at $x=1$ and $y=1200$ and rises up towards $2600$ at $x=3$ and falls sharply until $x=10$ with a long almost flat tail distribution until cell $300$.}
    	\caption{Position of First Navigable Table in Notebooks}
    	\label{fig:table_notebook_first_position}
    \end{subfigure}
    \caption{Presence of first navigable heading and table elements in the Notebooks. \textbf{Left (\ref{fig:hlevel_heatmap}):} shows the cell position of the first heading element and its level ($x$ axis) from the 1st cell in the notebook to the 10th ($y$ axis). \textbf{Right (\ref{fig:table_notebook_first_position}):} shows the cell position ($x$ axis) of the first table present in notebooks ($y$ axis) with a navigable table (excluding notebooks with 0 tables).}
\end{figure*}

To assess navigability, we calculate where the \textit{first} heading element lies in each notebook, as well as what type of heading it is (H1 through H6). In ~\Cref{fig:hlevel_heatmap}, the left most column (from top to bottom) shows the number of notebooks with an H1 in cell 1, cell 2, and so on down to cell 10. A majority  (28414 (28.90\%) notebooks) have a heading level 1 (\texttt{H1}) in the first cell. We speculate IDE integrations, JupyterLab, and Google Colab's default behavior, which adds an H1 in the first cell automatically,  may have resulted in this accessibility advantage.  As the first occurrence of a heading moves further away from the first cell indicated in \Cref{fig:hlevel_heatmap}, it is likely that screen reader users will skip a number of useful cells in the notebook. Screen reader users may also be confused by or skip cells when authors use a different heading level for a first heading  in cell 1, breaking the typically expected structure of the notebook or web based interactions. In \Cref{fig:hlevel_heatmap}, we see that 10955 (11.14\%) notebooks start with an H2 (row 1, column 2), 5728 (5.82\%) with H3 and so on down to H6 (.09\%). 47543 (48.36\%) notebooks contain a heading of \textit{any} level in the first cell of the notebook and 59.67\% (28414) of them correctly match our expectations to have a heading level 1 in the first cell. The high percentage of notebooks containing a heading indicate the high potential for improving accessibility of notebooks through improved navigability for screen reader users. The subsequent rows indicate the number of useful code cells potentially skipped by the screen reader navigation and show the frequency of occurrences of the first heading type in different cells of the notebooks. \Cref{fig:hlevel_heatmap} presents only a partial slice of the overall occurrence of heading in the notebooks, the detailed result is presented in \Cref{appendix:cell_distance_first_interactive}.

\subsubsection{Finding landmarks: Tables}
\label{sssec:tables}

We found that \numberOfNotebooksWithTables (\percentageNotebooksWithTables) of the notebooks have tables in them. Among these, a notebook contains  \meanNumberOfTablesInNotebooks tables on average, with \medianNumberOfTablesInNotebooks tables at the median, and the top 1 percentile of notebooks containing at least \lastPercentileNumberOfTables tables as shown in \Cref{fig:tablesInNotebooks}. The lack of tables in \numberOfNotebooksWithNoTables (\percentageNotebooksWithNoTables) of notebooks, consequently reduces the accessibility of notebooks and the possibility of glancing at data. 

The first occurrence of tables is typically the third cell in the notebook as shown in \Cref{fig:table_notebook_first_position}, perhaps because the first two cells are typically used for importing the necessary modules, and loading the required datasets for further analysis. This raises a further optimistic aspect of our metric -- such tables may be primarily present to check that data loaded correctly, rather than highlighting results after analysis is complete, limiting their utility for screen reader users to understand a notebook.

\subsection{Tooling Impacts on Notebook Accessibility}
\label{subsec:infrastructure_impact}

The accessibility of published notebooks on BVI consumers is impacted not only by authorship, but also by the tools used to export them. Jupyter notebooks allow the developers to export the notebook into various formats such as PDF, \LaTeX\ or HTML, among many others. The HTML format is widely used by notebook authors when releasing their notebooks because of the ease of sharing contents over the web. Many popular code hosting repository tools like GitHub or GitLab use the submitted raw \texttt{ipynb} notebook format files and convert them by exporting them into HTML when navigating to the file using a web browser. Thus we focus our analysis on HTML renderings. We use a mix of manual testing and automated accessibility testing tools to assess this. We measure:  

\begin{description}[style=unboxed,leftmargin=0cm,itemsep=1em]
    \item[Reachability of structural and graphical information:] As mentioned in \Cref{ssec:glanceability}, the ability to navigate to structural elements is important to accessibility. While our previous analysis looked at whether semantic information is  properly included in notebooks by authors (\Cref{sssec:headings} and \Cref{sssec:tables}), here we test the same question once notebooks have been rendered. To assess this, we performed manual screen reader testing, the only metric that we tested at small scale (only for 10 notebooks). This is an optimistic estimate because we did not test whether the landmarks were useful, only whether a screen reader user could get to them. We find significant concerns with the scalability of notebooks for screen reader users affecting navigability in 6/10 cases and present the detailed results in \Cref{ssec:reachability}.
    \item[Impact of Theme Choice on Accessibility  (\Cref{sssec:manualGlanceability}):] This metric captures  the accessibility impact due to the choice of theme when generating a notebook. Many themes are also used during authoring, so this may also impact authorship. This metric is optimistic since it may not capture all errors introduced by a theme due to the use of automated testing, which is known to be an optimistic measure of website accessibility~\cite{MankoffFT05}. The best theme we tested differed from the worst by  84.95\% overall; however we found that different themes performed differently on different accessibility testers and raised different types of errors within those testers. 

\end{description}

\subsubsection{Reachability of structural and graphical information}
\label{ssec:reachability}

\begin{table*}[!ht]
    \caption{Accessibility evaluation on randomly chosen notebooks of variable sizes (in bytes) ordered by percentile rank (P10-P100). For consistency we chose the \textit{light} theme in our evaluations. Only VoiceOver is shown for Safari, since the browser is no longer actively supported on Windows.
	$\checkmark$ indicates notebooks which pass the accessibility evaluation for glanceability, 
	$\ominus$ represents those which are functional, but not fully glanceable,
	\xmark\ represents those which fail glanceability evaluation for multiple reasons, 
	and $\perp$ represent the notebook which cause complete crashes of screen readers or browser tabs.}
% \smaller
	\begin{adjustbox}{width=.95\linewidth,center}
		\begin{tabular}{|l|l|l|lc|llc|lc|c|}
			\hline
			 &
			   &
			   &
			  \multicolumn{2}{c|}{\textbf{Microsoft Edge}} &
			  \multicolumn{3}{c|}{\textbf{Google Chrome}} &
			  \multicolumn{2}{c|}{\textbf{Firefox Nightly}} &
			  \textbf{Apple Safari} \\ \cline{4-11} &
			  \multirow{-2}{*}{\textbf{Rank}} &
			  \multirow{-2}{*}{\textbf{Size (Bytes)}} &
			  \multicolumn{1}{c|}{NVDA} &
			  \multicolumn{1}{c|}{JAWS} &
			  \multicolumn{1}{c|}{NVDA} &
			  \multicolumn{1}{c|}{JAWS} &
			  VoiceOver &
			  \multicolumn{1}{c|}{NVDA} &
			  \multicolumn{1}{c|}{JAWS} &
			  VoiceOver \\ \hline
			N1 &
			  1 (P10) &
			  630352 &
			  \multicolumn{1}{c|}{\cellcolor[HTML]{90EE90}{\checkmark}} &
			  \multicolumn{1}{c|}{\cellcolor[HTML]{90EE90}{\checkmark}} &
			  \multicolumn{1}{c|}{\cellcolor[HTML]{90EE90}{\checkmark}} &
			  \multicolumn{1}{c|}{\cellcolor[HTML]{90EE90}{\checkmark}} &
			  \cellcolor[HTML]{90EE90}{\checkmark} &
			  \multicolumn{1}{c|}{\cellcolor[HTML]{90EE90}{\checkmark}} &
			  \multicolumn{1}{c|}{\cellcolor[HTML]{90EE90}{\checkmark}} &
			  \cellcolor[HTML]{90EE90}{\checkmark} \\ \cline{1-11}
			N2 &
			  2 (P25) &
			  634340 &
			  \multicolumn{1}{c|}{\cellcolor[HTML]{90EE90}{\checkmark}} &
			  \multicolumn{1}{c|}{\cellcolor[HTML]{90EE90}{\checkmark}} &
			  \multicolumn{1}{c|}{\cellcolor[HTML]{90EE90}{\checkmark}} &
			  \multicolumn{1}{c|}{\cellcolor[HTML]{90EE90}{\checkmark}} &
			  \cellcolor[HTML]{90EE90}{\checkmark} &
			  \multicolumn{1}{c|}{\cellcolor[HTML]{90EE90}{\checkmark}} &
			  \multicolumn{1}{c|}{\cellcolor[HTML]{90EE90}{\checkmark}} &
			  \cellcolor[HTML]{90EE90}{\checkmark} \\ \cline{1-11}
			N3 &
			  3 (P50) &
			  721056 &
			  \multicolumn{1}{c|}{\cellcolor[HTML]{FFFFE0}{$\ominus$}} &
			  \multicolumn{1}{c|}{\cellcolor[HTML]{FFFFE0}{$\ominus$}} &
			  \multicolumn{1}{c|}{\cellcolor[HTML]{FFFFE0}{$\ominus$}} &
			  \multicolumn{1}{c|}{\cellcolor[HTML]{FFFFE0}{$\ominus$}} &
			  \cellcolor[HTML]{90EE90}{\checkmark} &
			  \multicolumn{1}{c|}{\cellcolor[HTML]{FFFFE0}{$\ominus$}} &
			  \multicolumn{1}{c|}{\cellcolor[HTML]{FFFFE0}{$\ominus$}} &
			  \cellcolor[HTML]{90EE90}{\checkmark} \\ \cline{1-11}
			N4 &
			  4 (P75) &
			  997861 &
			  \multicolumn{1}{c|}{\cellcolor[HTML]{FFFFE0}{$\ominus$}} &
			  \multicolumn{1}{c|}{\cellcolor[HTML]{FFFFE0}{$\ominus$}} &
			  \multicolumn{1}{c|}{\cellcolor[HTML]{FFFFE0}{$\ominus$}} &
			  \multicolumn{1}{c|}{\cellcolor[HTML]{FFFFE0}{$\ominus$}} &
			  \cellcolor[HTML]{90EE90}{\checkmark} &
			  \multicolumn{1}{c|}{\cellcolor[HTML]{FFFFE0}{$\ominus$}} &
			  \multicolumn{1}{c|}{\cellcolor[HTML]{FFFFE0}{$\ominus$}} &
			  \cellcolor[HTML]{90EE90}{\checkmark} \\ \cline{1-11}
			N5 &
			  5 (P85) &
			  1362512 &
			  \multicolumn{1}{c|}{\cellcolor[HTML]{90EE90}{\checkmark}} &
			  \multicolumn{1}{c|}{\cellcolor[HTML]{90EE90}{\checkmark}} &
			  \multicolumn{1}{c|}{\cellcolor[HTML]{90EE90}{\checkmark}} &
			  \multicolumn{1}{c|}{\cellcolor[HTML]{90EE90}{\checkmark}} &
			  \cellcolor[HTML]{90EE90}{\checkmark} &
			  \multicolumn{1}{c|}{\cellcolor[HTML]{FFFFE0}{$\ominus$}} &
			  \multicolumn{1}{c|}{\cellcolor[HTML]{FFFFE0}{$\ominus$}} &
			  \cellcolor[HTML]{90EE90}{\checkmark} \\ \cline{1-11}
			N6 &
			  6 (P90) &
			  1900465 &
			  \multicolumn{1}{c|}{\cellcolor[HTML]{FFFFE0}{$\ominus$}} &
			  \multicolumn{1}{c|}{\cellcolor[HTML]{FFFFE0}{$\ominus$}} &
			  \multicolumn{1}{c|}{\cellcolor[HTML]{FFFFE0}{$\ominus$}} &
			  \multicolumn{1}{c|}{\cellcolor[HTML]{FFFFE0}{$\ominus$}} &
			  \cellcolor[HTML]{90EE90}{\checkmark} &
			  \multicolumn{1}{c|}{\cellcolor[HTML]{FFFFE0}{$\ominus$}} &
			  \multicolumn{1}{c|}{\cellcolor[HTML]{FFFFE0}{$\ominus$}} &
			  \cellcolor[HTML]{90EE90}{\checkmark} \\ \cline{1-11}
			N7 &
			  7 (P95) &
			  1915381 &
			  \multicolumn{1}{c|}{\cellcolor[HTML]{FFFFE0}{$\ominus$}} &
			  \multicolumn{1}{c|}{\cellcolor[HTML]{FFFFE0}{$\ominus$}} &
			  \multicolumn{1}{c|}{\cellcolor[HTML]{FFFFE0}{$\ominus$}} &
			  \multicolumn{1}{c|}{\cellcolor[HTML]{FFFFE0}{$\ominus$}} &
			  \cellcolor[HTML]{90EE90}{\checkmark} &
			  \multicolumn{1}{c|}{\cellcolor[HTML]{FFFFE0}{$\ominus$}} &
			  \multicolumn{1}{c|}{\cellcolor[HTML]{FFFFE0}{$\ominus$}} &
			  \cellcolor[HTML]{90EE90}{\checkmark} \\ \cline{1-11} 
			N8 &
			  8 (P99) &
			  10955553 &
			  \multicolumn{1}{c|}{\cellcolor[HTML]{FFCCCB}{\xmark}} &
			  \multicolumn{1}{c|}{\cellcolor[HTML]{FFCCCB}{\xmark}} &
			  \multicolumn{1}{c|}{\cellcolor[HTML]{FFCCCB}{\xmark}} &
			  \multicolumn{1}{c|}{\cellcolor[HTML]{FFCCCB}{\xmark}} &
			  \cellcolor[HTML]{90EE90}{\checkmark} &
			  \multicolumn{1}{c|}{\cellcolor[HTML]{FFCCCB}{$\perp$}} &
			  \multicolumn{1}{c|}{\cellcolor[HTML]{FFCCCB}{$\perp$}} &
			  \cellcolor[HTML]{90EE90}{\checkmark} \\ \cline{1-11} 
			N9 &
			  9 (P100) &
			  103790428 &
			  \multicolumn{1}{c|}{\cellcolor[HTML]{FFCCCB}{$\perp$}} &
			  \multicolumn{1}{c|}{\cellcolor[HTML]{FFCCCB}{$\perp$}} &
			  \multicolumn{1}{c|}{\cellcolor[HTML]{FFCCCB}{$\perp$}} &
			  \multicolumn{1}{c|}{\cellcolor[HTML]{FFCCCB}{$\perp$}} &
			  \cellcolor[HTML]{90EE90}{\checkmark} &
			  \multicolumn{1}{c|}{\cellcolor[HTML]{FFCCCB}{$\perp$}} &
			  \multicolumn{1}{c|}{\cellcolor[HTML]{FFCCCB}{$\perp$}} &
			  \cellcolor[HTML]{90EE90}{\checkmark} \\ \hline
			\end{tabular}
	\end{adjustbox}
	\label{table:accessibility_evaluation_table}
\end{table*}

\Cref{table:accessibility_evaluation_table} shows the results of manual testing of the accessibility of notebooks. We sampled notebooks of a representative range of  sizes as we observed that large notebooks were crashing browsers during the exploratory phase of this research. Through the sampling based manual analysis, we identified that notebooks of large sizes can cause accessibility breakdowns, crashing screen readers and browsers on windows, indicated by $\perp$ in \Cref{table:accessibility_evaluation_table}. The results suggest that at least 1\% of all notebooks are fully inaccessible due to their large sizes and cause screen readers or web browser tabs to completely crash.
We speculate that the difference in how JAWS and NVDA read webpages through a virtual buffer \vs VoiceOver's ability to directly interact with the browser causes the difference in accessibility breakdowns visible in the table~\cite{rethinking2017marco}. Addressing these breakdowns in windows screen readers is critical due to their popularity among the BVI community.

Notebook N5 is an interesting case. Although other notebooks smaller than it fail, it passes for both NVDA and JAWS on Edge and Chrome. Firefox Nightly however was not able to find all the required headers or tables present in the notebook therefore marking its status as a functional but not fully glanceable notebook. We looked more deeply into this and found that one possible cause is that N5 did not contain any programmatically generated graphics. 

Additionally, our manual accessibility checks demonstrate that for \textit{any notebook size}, both JAWS and NVDA with Chrome, Edge and Firefox only detect the first programmatically generated graphic despite the existence of more than one in the subsequent cells. While navigating by graphic will jump focus to other images that are manually added to markdown cells, both JAWS and NVDA do not jump to any graphics that are encoded as base64 strings that occur in subsequent code cells, a significant accessibility problem (indicated by $\ominus$ in \Cref{table:accessibility_evaluation_table}). We further tested this behavior by adding synthetically generated base64 encoded images of different sizes into HTML outputs randomly chosen from our data, and to new notebooks with just two cells and observed similar behavior. It is important for screen reader users to be informed of the presence of  these images and to be able to navigate to them. Current attempts at navigation using popular NVDA and JAWS screen readers  completely skip images. VoiceOver on the Mac performs the best with no crashes and detects all images, headings, and tables -- satisfying the requirements for notebook glanceability we set forth (denoted by $\checkmark$), and enables screen reader users to navigate to them using the VoiceOver rotor -- a single-key navigation equivalent for VoiceOver. 
Though the notebooks are glanceable using VoiceOver on the Mac, it is critical that Windows based screen readers be able to do the same due to their wide  adoption among BVI users~\cite{webaim2021screenreaderuse}.

\subsubsection{Impact of Theme Choice on Accessibility}
\label{sssec:manualGlanceability}

\begin{figure*}[h!]
    \centering
   \begin{subfigure}[b]{.45\linewidth}
        \includegraphics[width=\linewidth]{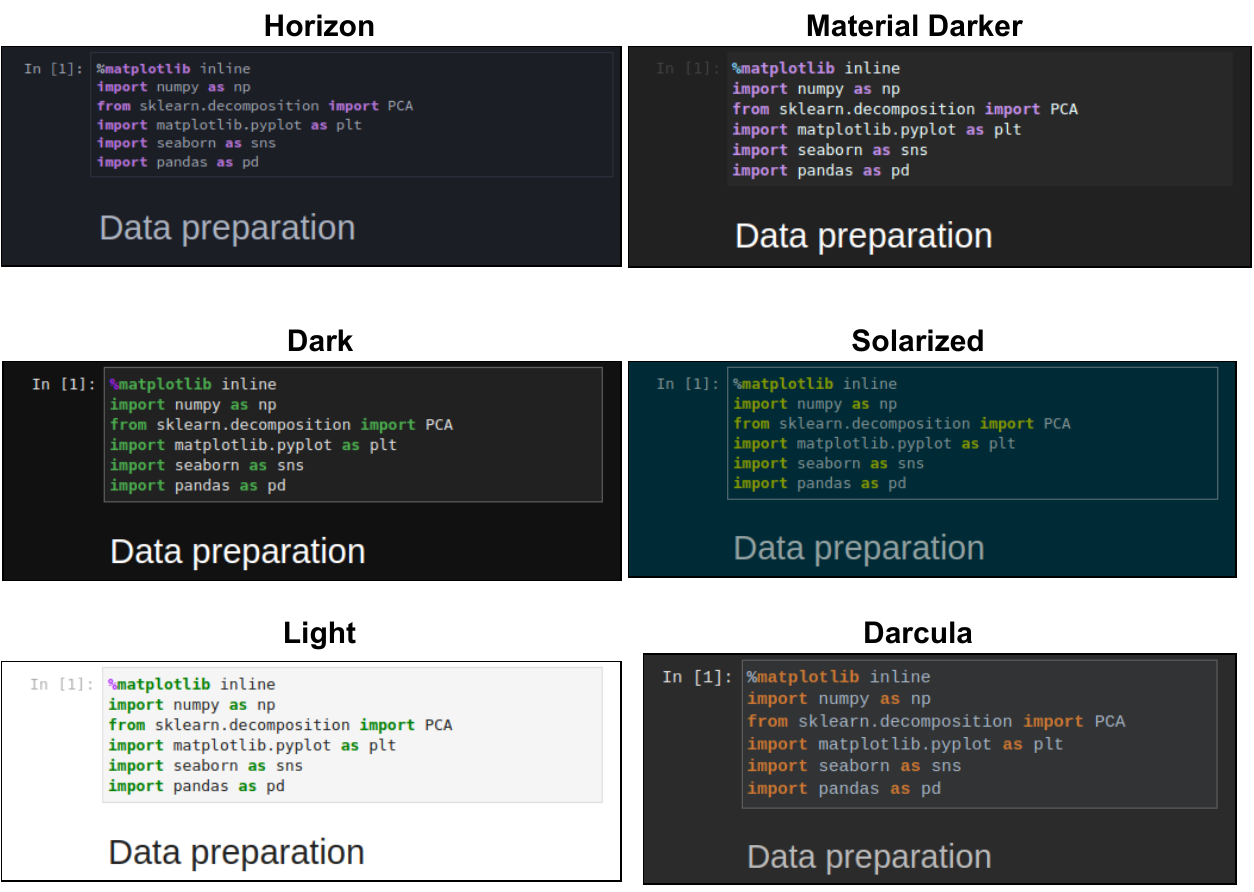}
        \Description[Examples of notebooks rendered in different Themes]{Figure shows a 2 x 3 grid of screenshots all containing the same code snippet from a notebook with varying themes applied. The top row contains horizon and material darker, the second row contains dark and solarized, and the last row contains light and darcula theme representations of notebook cells.}
        \caption{Examples of notebooks rendered in different Themes}
        \label{fig:notebook_themes_sample}
    \end{subfigure}
    \hfill
    \begin{subfigure}[b]{.45\linewidth}
        \includegraphics[width=\linewidth]{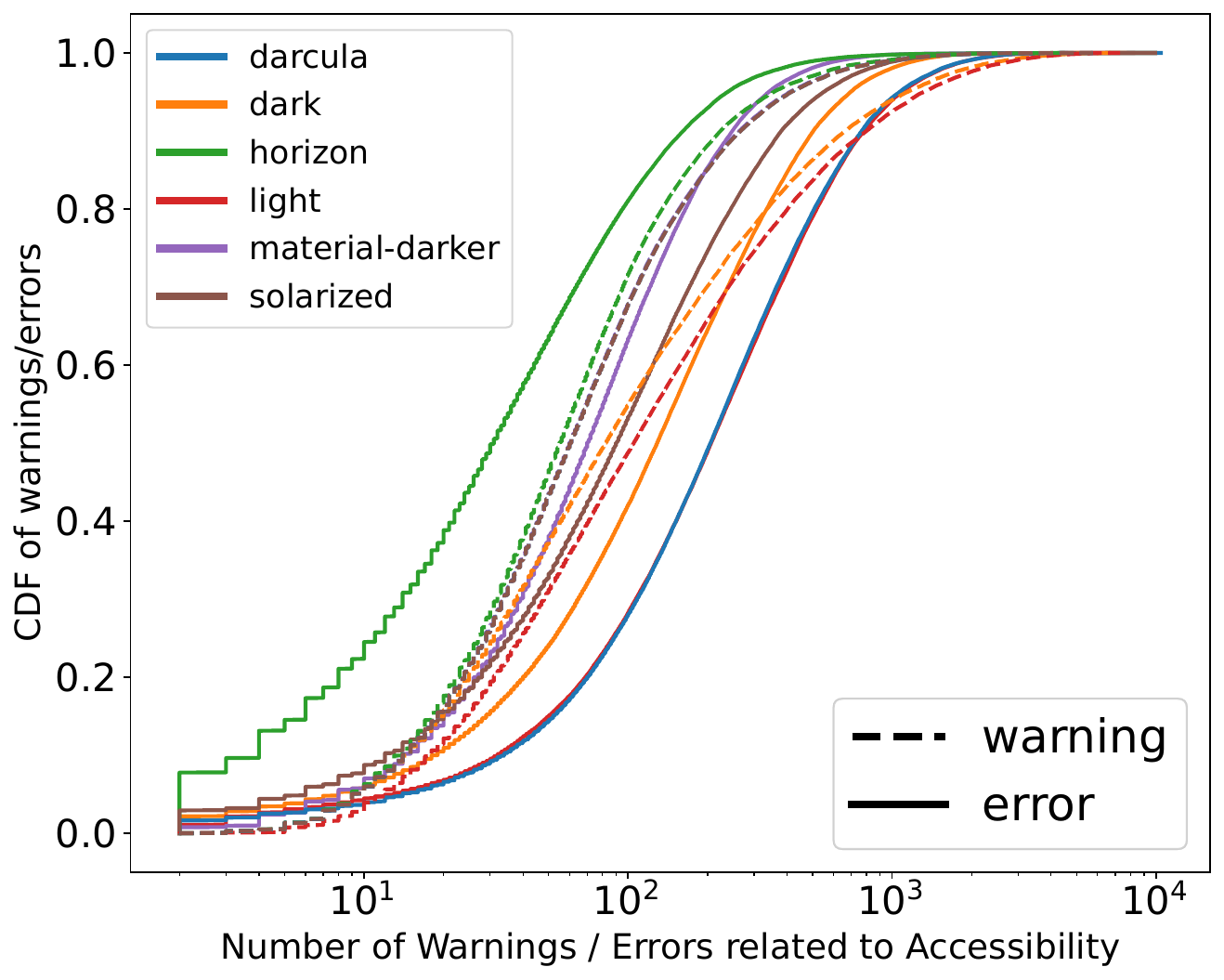}
        \Description[CDF of accessibility warnings and errors based on theme]{Figure indicates the CDF of the accessibility warnings and errors reported by axe and htmlcs on various themes in our experiments. The figure contains a total of 12 colored lines, using 6 distinct colors and two line styles. The solid lines represent the distribution of errors per theme, and the dashed line represents the number of warnings per theme. The solid blue and red lines indicating darcula and light are to the extreme right side of the lines in the figure indicating the most number of accessibility errors compared to other themes, the best performing theme is indicated by the green solid line representing horizon theme which indicates an 85\% improvement over the darcula and light themes. Other themes in our study lie between these two lines whose medians are at 31 and 160 respectively. The x axis is represented by a log scale indicator of the number of warnings / errors related to accessibility and the y axis indicates the CDF ranging from 0-1.}
        \caption{CDF of accessibility warnings and errors based on theme}
        \label{fig:accessibility_warnings_errors}
    \end{subfigure}

    \caption{Customizability of Notebook experiences and its accessibility implications. \textbf{Left (\ref{fig:notebook_themes_sample})}: shows the visual difference in notebooks when applying different themes in our evaluation.\textbf{Right (\ref{fig:accessibility_warnings_errors})}: shows the distribution of the number of accessibility errors and warnings reported by accessibility engines on the same set of notebooks exported into multiple themes. }
    \label{fig:theme-warnings-errors-and-example}
\end{figure*}

To understand the impact of the choice of theme to accessibility of the notebook, we evaluate each notebooks' accessibility using six themes, two of which are the default (light and dark), and the four others which are popular defaults used by various IDEs.  The use of color schemes to modify the visual style of the editor interface is a common practice and allows developers to improve their productivity due to their ability to make code easier to visually read and understand. These visual differences are  summarized for the six themes we selected in \Cref{fig:notebook_themes_sample}. Applying  color schemes such as high contrast themes on IDEs has been widely adopted by developers for both accessibility (low vision developers), and aesthetic reasons. Although Jupyter notebooks by default are exported to HTML using the  \textit{light} theme, notebook authors can  specify a different theme to use and export the notebook into.  However, as summarized in \Cref{fig:theme-warnings-errors-and-example}, these themes also differ significantly by the amount of warnings and errors related to accessibility that we found when we used automated testing on them.  
 
We found that the horizon theme, which is the default for the popular VSCode IDE, performs the best, with the fewest accessibility errors ($\mu$=67.52 ($\sigma$=138.97) errors). It was 84.95\% better than the default light theme provided by the Jupyter IDE, which had a mean of 335.40 ($\sigma$=393.72) errors. This difference is statistically significant according to the paired-samples t-test (\textit{t(95101)=198.05, p $<$ .001}). 
\Cref{fig:accessibility_warnings_errors} summarizes all six themes using a CDF showing the distribution of the number of errors (solid lines) and warnings (dashed lines) reported by the aXe and HTMLCS accessibility engines on the exported HTML versions of the notebooks.

\begin{table*}[h!]
    \caption{Description of Accessibility Errors found in the dataset and their impact on Users (obtained from aXe documentation). This table provides the key for errors in \Cref{fig:heatmap_errors} and specific  principles which are violated as per the WCAG2AA guidelines}.
	\begin{adjustbox}{width=\linewidth,center}
	\begin{tabular}{|l|l|p{3.5cm}|p{6.5cm}|l|}
		\hline
		\textbf{Axe} &
		\textbf{HTMLCS} &
		\textbf{WCAG 2AA} &
		\textbf{Error Description} &
		\textbf{Impact} \\ \hline
		AXE-E1 &
		HTMLCS-E1 &
		1.4.3 G18 Fail &
		Text elements must have sufficient color contrast against the background &
		\cellcolor[HTML]{FFCCC9}Serious \\ \hline
		AXE-E2 &
		HTMLCS-E2 &
            1.1.1 H37 &
		Images must have alternate text &
		\cellcolor[HTML]{FD6864}Critical \\ \hline
		AXE-E3 &
		HTMLCS-E7 &
            1.4.3 F24 &
		Links must be distinguished from surrounding text in a way that does not rely on color &
		\cellcolor[HTML]{FFCCC9}Serious \\ \hline
		AXE-E4 &
		\cellcolor[HTML]{C0C0C0} &
            2.4.4 H77, H78, H79, H80, H81, H33 &
		Links must have discernible text &
		\cellcolor[HTML]{FFCCC9}Serious \\ \hline
		AXE-E5 &
		\cellcolor[HTML]{C0C0C0} &
            2.4.1 G1, G123, G124, H69 &
		Page must have means to bypass repeated blocks &
		\cellcolor[HTML]{FFCCC9}Serious \\ \hline
		AXE-E6 &
		\cellcolor[HTML]{C0C0C0} &
            1.2.1 G159, G166 &
		\textless{}audio\textgreater elements must have a captions \textless{}track\textgreater{} &
		\cellcolor[HTML]{FD6864}Critical \\ \hline
		AXE-E7 &
		\cellcolor[HTML]{C0C0C0} &
		&
		aria-hidden elements do not contain focusable elements &
		\cellcolor[HTML]{FFCCC9}Serious \\ \hline
		AXE-E8 &
		\cellcolor[HTML]{C0C0C0} &
		&
		ARIA input fields must have an accessible name &
		\cellcolor[HTML]{FFCCC9}Serious \\ \hline
		AXE-E9 &
		\cellcolor[HTML]{C0C0C0} &
		&
		Certain ARIA roles must be contained by particular parent elements &
		\cellcolor[HTML]{FD6864}Critical \\ \hline
		AXE-E10 &
		\cellcolor[HTML]{C0C0C0} &
		&
		All page content must be contained by landmarks &
		\cellcolor[HTML]{FFFC9E}Moderate  \\ \hline
		\cellcolor[HTML]{C0C0C0} &
		HTMLCS-E3 &
            4.1.1 F77 &
		Duplicate ID attribute value found on the web page. &
		\cellcolor[HTML]{CBCEFB}Minor \\ \hline
		\cellcolor[HTML]{C0C0C0} &
		HTMLCS-E4 &
            1.3.1 H43, H63 &
		Tables not using header or scope attributes &
		\cellcolor[HTML]{FD6864}Critical  \\ \hline
		\cellcolor[HTML]{C0C0C0} &
		HTMLCS-E5 &
            1.3.1 H43 Headers Required &
		Table Header Required and currently not used &
		\cellcolor[HTML]{FD6864}Critical \\ \hline
		\cellcolor[HTML]{C0C0C0} &
		HTMLCS-E6 &
            4.1.2 H91 A EmptyNoId &
		Anchor Elements with no ID or link content &
		\cellcolor[HTML]{FFCCC9}Serious \\ \hline
		\cellcolor[HTML]{C0C0C0} &
		HTMLCS-E8 &
            4.1.2.H91 A NoContent &
		Anchor Elements with Valid Link but no link text content &
		\cellcolor[HTML]{FFCCC9}Serious \\ \hline
		\cellcolor[HTML]{C0C0C0} &
		HTMLCS-E9 &
		4.1.2 H91 Button Name, 4.1.2 H91 {[}Element{]} Name &
		This {[}Element type{]} does not have a name available to an accessibility API. &
		\cellcolor[HTML]{FD6864}Critical \\ \hline
	\end{tabular}
	\end{adjustbox}
	\label{table:accessibility_errors_table}
\end{table*}

\begin{figure*}
	\begin{subfigure}[b]{.45\linewidth}
		\includegraphics[width=\linewidth]{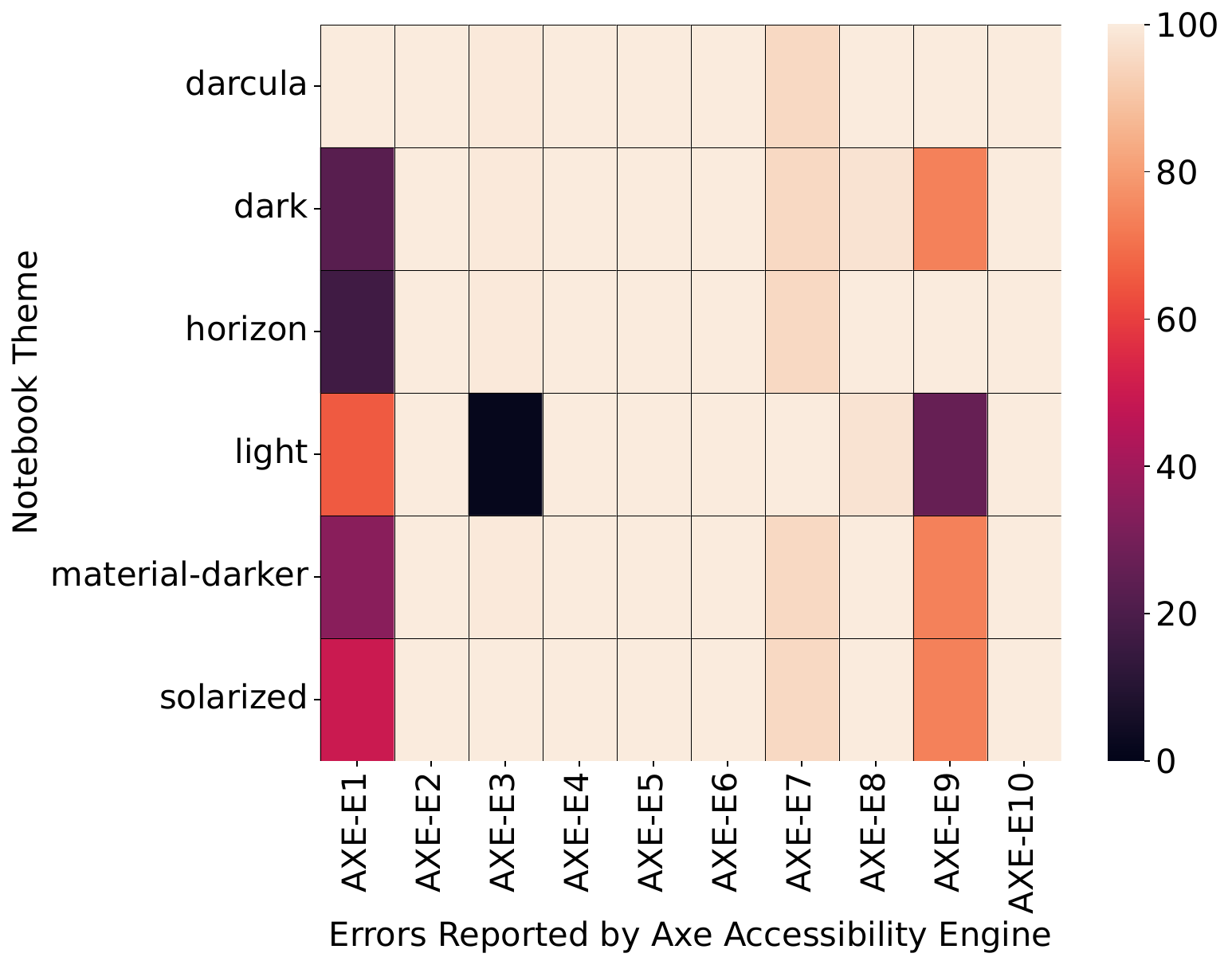}
		\caption{Heatmap of Errors Types from aXe Engine}
		\label{fig:heatmap_axe_errors}
	\end{subfigure}
	\hfill
	\begin{subfigure}[b]{.45\linewidth}
		\includegraphics[width=\linewidth]{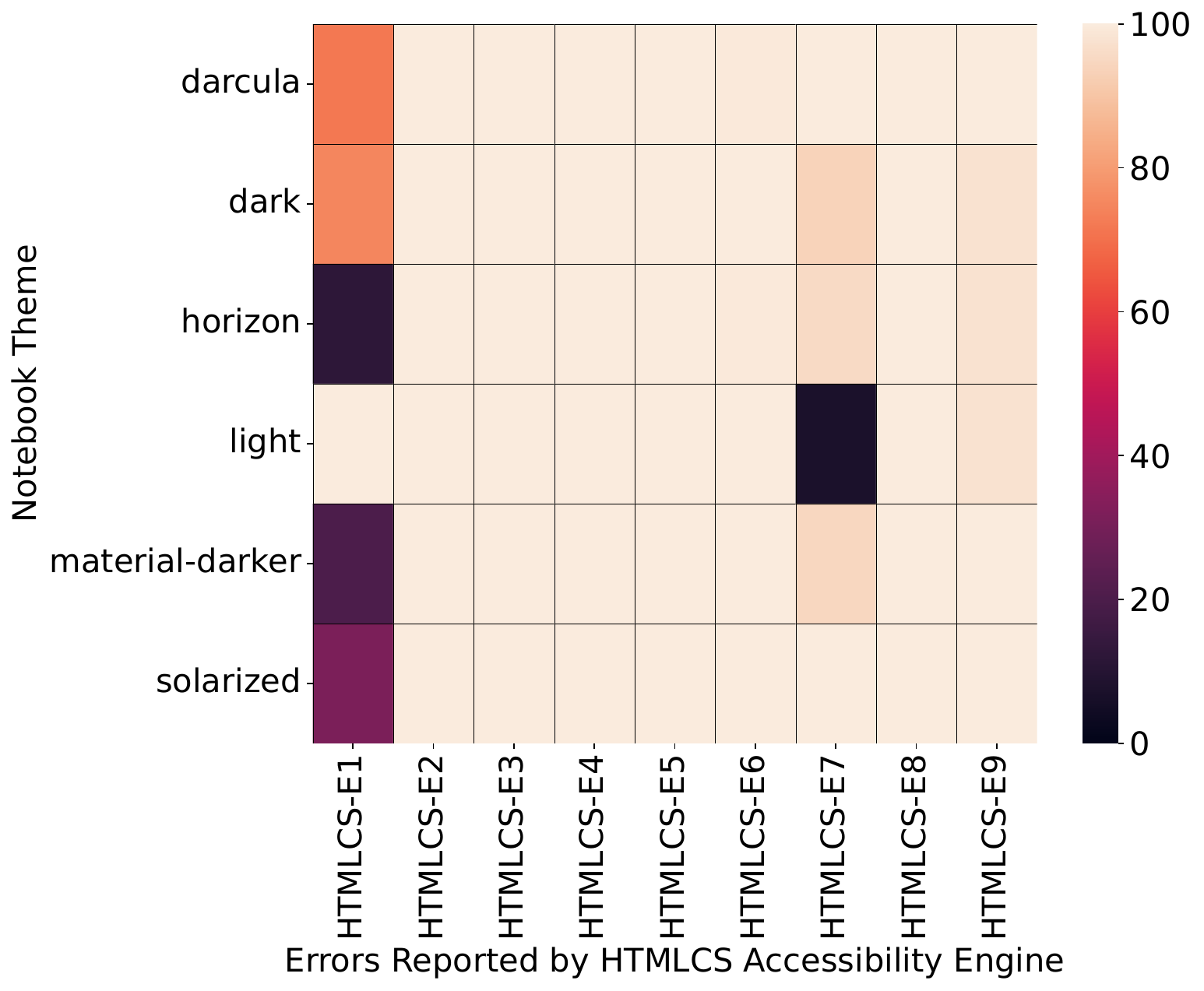}
		\caption{Heatmap of Errors Types from HTMLCS Engine}
		\label{fig:heatmap_html_errors}
	\end{subfigure}
        \Description[Heatmap of Error Types]{Figure contains two subfigures, both representing the relative difference in the errors found between themes by each accessibility engine. Axe is represented by the figure on the left and htmlcs is represented by the figure on the right. The heatmap of errors for axe in the left subfigure indicate the percentage of errors of different types. The type of errors are presented on the x axis and the themes along the y axis with each cell ranging between 0 and 100.0\%. The corresponding legend of the color bar is to the right side of the figure. The subfigure on the right is similar to the one on the left, but contains 9 error types instead of 10 because of a different engine type. The key for the error codes along the x axis for both figures are presented in Table 4.}
	\caption{Relative difference in errors found between themes by each accessibility engine. \textbf{Left (\ref{fig:heatmap_axe_errors})} shows the errors reported by the aXe engine, \textbf{Right (\ref{fig:heatmap_html_errors})} shows the errors reported by the HTMLCS engine.}
	\label{fig:heatmap_errors}
\end{figure*}

Looking more closely at error type can help us to explore the range of ways in which notebook exports affect accessibility.
We analyze the results of our accessibility scans and identify 10 error categories reported by the aXe engine, and 9 error categories reported by the HTML Code Sniffer  (HTMLCS) engine whose total occurances across all the notebooks differ based on the theme they were run on. In \Cref{fig:heatmap_errors} we present the relative heatmap of these errors by comparing the number of errors in each theme to the maximum number of errors reported for each error type. We present the details of the error code, and its impact on user accessibility enumerated through AXE-[1-10] and HTMLCS-[1-9] in \Cref{table:accessibility_errors_table}.

While our results from \Cref{fig:accessibility_warnings_errors} presented in \Cref{sssec:manualGlanceability} indicate that a change in theme would significantly improve the accessibility of the notebooks, the results in \Cref{table:accessibility_errors_table} indicate some unintended consequences of theme changes and also indicate the impact of the tools chosen to test accessibility. Of the 16 unique types of errors which differ among themes, both accessibility engines agree on only three which are grouped together due to their similarity (AXE-E1$\sim$HTMLCS-E1, AXE-E2$\sim$HTMLCS-E2, AXE-E3$\sim$HTMLCS-E7). The aXe scanner assiduously  identifies seven other issues as errors, some of which are typically considered warnings or notices by other tools or in the WCAG2AA standard  specification~\cite{wcag2aa}. HTMLCS identifies six other accessibility errors which are not identified by aXe. Together, the two engines uncover 6 critical accessibility error categories, eight serious ones, and one moderate and minor error category respectively. Our findings demonstrate  the value of running multiple accessibility engines while evaluating for accessibility challenges when employing automated mechanisms.

As is visible in \Cref{fig:heatmap_errors}, there are some disagreements between the accessibility tools. For example,  the   color contrast  metric result in aXe ranks the darcula theme as most inaccessible, while HTMLCS  considers the light theme to be the most inaccessible (AXE-E1$\sim$HTMLCS-E1). Similarly,  despite  being considered inaccessible due to color contrast issues with the background, the light theme performs the best when addressing the challenge of link distinguishability -- as indicated by the dark black squares in the ``light'' row on the AXE-E3 column in \Cref{fig:heatmap_axe_errors}, and HTMLCS-E7  column in \Cref{fig:heatmap_html_errors}. The difference in errors due to \texttt{aria} attributes such as \texttt{aria-hidden}, and \texttt{aria-parents} are also striking (AXE-E9). We manually inspected 5 notebooks (of the 38 notebooks which are affected) with the most AXE-E9 errors and found problems such as an out of date version of MathJax, incorrect  \LaTeX\ notations for mathematical expressions, and programmatic inclusion of stale documentation.  Interestingly, these are all sources outside the computational notebook ecosystem and introduced during authoring.

\section{Discussion and Recommendations}
\label{sec:discussion}

Our work is the first large scale analysis of computational notebooks done from an accessibility perspective. Our study of 100,000 notebooks provide insights not only about the current state of accessibility of notebooks,  but also suggest directions for future research. We note that our analysis does not directly address the fact that web-based notebook editors are not accessible to BLV authors. We do not say more about this as it was not a focus of our data analysis, but it is an important domain for future work to address~\cite{manfromjupyter2020accessibility}.

\subsection{Improving Artifact Accessibility}
\label{subsection:improving_artifact_accessibility}

We found that notebook images almost universally lack ALT text, and are usually created with \texttt{matplotlib} or a library built on top of it (\texttt{seaborn}). We also found a lack of accessible, tabular information associated with figures. Many of these concerns could be improved upon by providing additional options to developers using the libraries or minor changes to the tool defaults.

\subsubsection{Including alternate text in images}
\label{subsubsection:including_alt_text_in_images}

During the programmatic creation of tables and images, there is a great deal of available knowledge that could be used to improve their accessibility. First, it is possible to check for the presence of axis labels, validate color contrast, and even generate basic ALT text that would be significantly better than what we found (\Cref{fig:alt-text-word-cloud}), all based on information available when the chart is instantiated. For example, \texttt{matplotlib} during the generation of a line chart could include alt-text information indicating the type of chart, the color and number of lines, and the labels along various axis. Proprietary statistics software such as SAS/STAT include such metadata information in the graphics generated based on the context such as the function call, or the data being visualized~\cite{SASAltDescription}. Second, interactive support in notebooks could be designed to help users encode image descriptions according to Lungard \etals four-level semantic model~\cite{lundgard2021accessible}. Third, it is very feasible to modify a tool such as \texttt{matplotlib}, the most used plotting tool as observed in our analysis (presented in \Cref{table:top-10-function-calls}), and provide programmatic methods to embed ALT text in  PNG images (the most common image type it produces in our data (\Cref{table:top5-output-types})). 
ALT text could be provided manually by the notebook authors, similar to the ability for web developers to customize the \texttt{alt} attribute in HTML \texttt{<img>} tags to describe images. 

To demonstrate the feasibility of our suggestion, we updated the \texttt{matplotlib} backend for \texttt{PNG} formats, by modifying the modules' source code (\texttt{image.py}, \texttt{backend\_(png$|$pdf).py}), to use the standardized Exchangeable Image File Format (\texttt{EXIF}) and include alt text information which could be passed as image metadata by making less than 10 lines of code changes to the open source code. Such image description information can be encoded in a serialized byte format  against the \texttt{0x010e (270)} byte delimiter following the EXIF standards. Following the established and existing standards would allow consumers of notebook formats, such as Jupyter, VSCode, and various IDEs to embed ALT text into the figures and make it available to screen reader users. We further verified that the descriptions included using our proposed alt argument can be included by existing tools like \texttt{nbconvert} and included in the HTML conversions of the notebook.  

Our artifacts released as a part of this work utilize these changes and include ALT metadata description in the resulting images. We leave their ability to be automatically recognized by notebook software for future open source efforts motivated by this research.

\subsubsection{Leveraging Interactivity on the Web}
\label{subsubsection:leveraging_interactivity}

Visualizations on web pages (using libraries like d3.js, highcharts, jQuery charts), are often generated during the page load event establishing a separation between the data and the functions responsible for creating and rendering the visualizations, enabling interactivity, and making it possible for extensions to build additional accessibility features~\cite{Sharif:2022:VoxLens}. 

Future work could explore comprehensive API designs similar to those that are provided by Apple to sonify charts~\cite{appleaudiographs}, to replace static base64 encoded images returned from the Jupyter python kernels' message passing interface with a representation of only the data required offloading the visualization capabilities to the Jupyter front end. This could allow web exports of notebooks to separate the data and functions responsible for visualizations, thus making it feasible for including \textit{accessible}, and interactive graphs.

\subsubsection{Presenting multiple data representations}
\label{subsubsection:presenting_multiple_data_representations}

Similar innovations could be added to the tools used in notebooks to improve table prevalence and accessibility. For example, we found that tables don't always accompany charts, despite this being a best practice \cite{Elavsky:2022:Chartability}. 
It may be possible to extend modules such as matplotlib, or seaborn to return a table representing the various data points in the image being created, or for notebook tools to perform code analysis of popular data processing libraries like pandas (\Cref{table:top-10-function-calls}) and automatically insert the intermediate data representations passed to the visualization functions as a table. For example, Notebooks with the understanding of a code cell using a \texttt{pandas} dataframe in a variable `\texttt{df}' calling the \texttt{lineplot(data=df, x=`label', y=`data')} could create a snapshot of the data corresponding to the `label' and `data' columns in its metadata and present a table along with the resulting figure. Future accessibility efforts to improve this understanding could develop targeted heuristics to assess the relevance of tables and other data representations in notebooks.

\subsubsection{Addressing Large Tables}
\label{subsubsection:addressing_large_tables}

Our characterization of tables used in computational notebook software shows that tables used in notebooks are typically very wide (presented in \Cref{fig:numberOfRowsColumnsInTables}). Additionally, Wang~\etal{} highlight table navigation challenges experienced by blind web users, finding that screen reader users have difficulty keeping track of context when navigating long tables~\cite{Wang:2022:webDataTables}. Programmatic support could be designed to ensure that the defaults for representing tabular data are as accessible as possible, though additional research is needed to understand  the best way to accessibly represent tables for data sets with large numbers of columns. Extensive work has been done by a few contributors to the Notebooks4All effort to provide guidance on making table outputs accessible~\cite{dfA11y}.

Wang~\etal{}~\cite{Wang:2022:webDataTables} also find that screen reader users encounter incorrectly marked up tables. Though not significant, our accessibility scan results do indicate the presence of table related errors, possibly generated by data processing libraries such as pandas when specific constants to improve their accessibility are not set~\cite{dfA11y}.  Code analyses extending the one described in our pipeline (\Cref{subsec:data_processing_pipeline}), combined with accessibility scans of programmatically generated tables with  exhaustive combinations of parameters that have an impact on table output, could help identify root causes and target accessibility improvements. 

\subsection{Accessible Notebook Authoring}
\label{subsection:improving_authoring}

We found that many notebooks have an H1 in their first cell, but consistent and proper use of headers is by no means universal. Further, only a third of notebooks have even one table in them. Improving these authoring practices could improve notebook accessibility.

Computational notebook tools could be updated to provide notices to the authors about incorrect usage of headings when violating the heading order or existence in the WCAG2 guidelines. Similarly, they could suggest places where authors may want to summarize what has happened so far in a table. 
This could be valuable for all notebook authors, regardless of disability, since tables convey valuable information about the structure of the data being processed in the subsequent cells of the  notebook . These practices, along with increased figure accessibility, can improve notebook comprehensibility and glanceability. Future efforts could focus on understanding and improve notebook glanceability to further explore what glanceability means to a BVI user based on the task at hand (authoring \vs consuming) and incorporate these best practices into tools such as linters for Jupyter Notebooks, to support accessible notebook authoring.

\subsection{Accessible Notebook Consumption}
\label{subsection:improvements_to_notebook_infrastructure}

Notebooks that are converted to HTML may be accessible to consumers if that HTML is accessible. When we used the most accessible theme (Horizon, which is not the default that most authors are likely to use), we found a mean of 68 accessibility errors, when testing notebooks with automated tools. This was significantly better than the worst theme, which had an average of 335 errors.

These results are particularly striking since they imply that changing the default theme of the notebook software during export, or enabling the ability to toggle themes in the exported notebooks, could significantly improve the overall accessibility. While changing the theme is not the remedy to all accessibility issues in notebooks, it is an important factor to consider when improving and addressing the challenges of notebook accessibility.
 
As a first step, the Jupyter community could explore conducting an accessibility evaluation of their theme(s), address existing accessibility issues, and provide improved accessible defaults.  
Future explorations could investigate the feasibility of integrating accessibility checkers such as aXe, or HTMLCS into the \texttt{nbconvert} process or within the notebook programming environments. 
Notebook authors hosting their notebooks on source code repository systems such as GitHub could also consider integrating accessibility engines to test the export of notebooks into their continuous integration and deployment pipelines (CI/CD) which could highlight and prevent accessibility issues during notebook creation~\cite{CIA11y}.

\subsection{Reflections on Research Methodology}
\label{subsection:reflections_on_research_methodology}

Large scale characterization work does not frequently occur in accessibility research. Mack \etal{}, in their systematic literature survey of accessibility papers, report on methodology and find that less than 6\% of accessibility papers included large scale analyses~\cite{mack-chi21}.
Our work demonstrates the use of large scale data collection and characterization to understand the (in)accessibility of computational notebooks, and the sources for these errors from data that need not involve burdening people with disabilities. In choosing our approach, we were cognizant that not all accessibility issues are captured by  our approach, and thus we drilled down a few occurrences of accessibility issues to observe factors that may require targeted investigation. We hope that this methodology can be applied to other areas where accessibility breakdowns could happen without placing a burden on disabled stakeholders. To facilitate further research on accessibility of computational notebooks, we make our dataset and analysis scripts publicly available~\cite{Potluri_Notably_Inaccessible_2023} and link to them in IncluSet~\cite{Kacorri:2020:IncluSet} and Zenodo~\cite{potluri_venkatesh_2023_8185050}.

\section{Conclusion}
\label{section:conclusion}

We present the first large-scale analysis of computational notebooks aimed at gaining an understanding about their accessibility. Current accessibility work to make computational notebooks accessible are focused on remediating the accessibility of the interface used to create these notebooks. While this is important, we identify that  programming environments,  authoring practices and distribution mechanisms chosen by authors can also have an impact on the accessibility of computational notebooks. 
We find that the data artifacts presented in notebooks cause inaccessibility, and tools to create these artifacts some times do not have mechanisms to make data accessible. 
We find that notebook interfaces can use web semantics to make data and notebooks glanceable, although they are not frequently used by authors. Finally, we observe that customizations used by authors can impact accessibility and potentially make notebook authoring and consumption more accessible. Finally, we make actionable recommendations  to increase the accessibility of data artifacts, tools, and notebook infrastructures.

% \input{proposal.tex}
% \input{related_work.tex}

\begin{acks}
We thank Richard Anderson and Kurtis Heimerl for the cloud infrastructure access and their thoughtful feedback  through the course of this work.  We thank Tim Althoff, Ken Gu, and Ashish Sharma from the Behavioral Data Science group for their initial feedback and comments related to this work. Executing this research would not have been possible without the gracious infrastructure support provided by the UW CSE Support team especially Stephen Spencer, and Aaron Timss. We would also like to thank Aaron Goldenthal for their initial work in improving \textit{pa11y} for continued support on newer operating systems which was further extended and instrumented to perform large scale accessibility analysis. We thank the members of University of Washington's Make4All and ICTD labs for their feedback and support through the course of this work. Venkatesh Potluri was supported by the Apple Scholars in AI/ML PhD fellowship. This work is supported by the \grantsponsor{GS01NSF}{National Science Foundation (NSF) Eng Diversity Activities (EDA)}{https://www.nsf.gov/awardsearch/showAward?AWD_ID=2009977}~\grantnum{GS01NSF}{2009977}, and the \grantsponsor{GS02CREATE}{Center for Research and Education on Accessible Technology and Experiences (CREATE)}{https://create.uw.edu/}.

\end{acks}

\bibliographystyle{ACM-Reference-Format}
\bibliography{references,vref}

\newpage

\appendix
\onecolumn

\section{Cell Distance to First Interactive Heading Element}
\label{appendix:cell_distance_first_interactive}

\begin{figure*}[h]
    \centering
    \includegraphics[width=\linewidth]{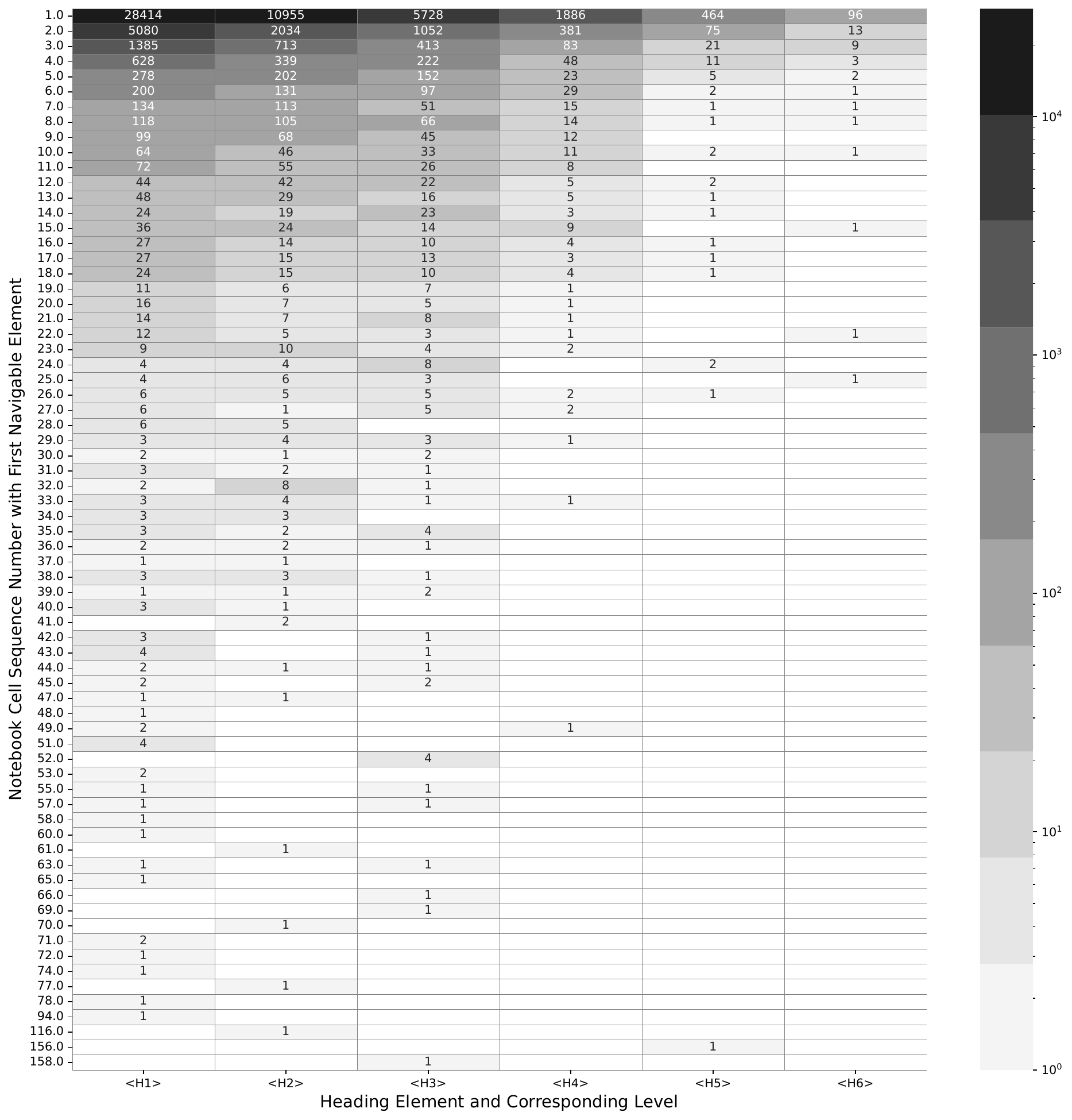}
    \caption{Complete heatmap used to derive the partial \Cref{fig:hlevel_heatmap} indicating the presence of first navigable heading and table elements in notebooks. Each cell in the figure shows the number of notebooks with the first occurrence of a heading level at a given code cell.}
    \label{fig:appendix_cell_distance_interactive_heading}
\end{figure*}

\newpage

\section{Types of Outputs in Notebooks}
\label{appendix:types_of_outputs_in_notebooks}

\begin{table*}[h]
    \caption{Complete List of Output Types in Notebooks extending the result presented in \Cref{table:top5-output-types}}
	\begin{adjustbox}{width=.7\linewidth,center}
    \begin{tabular}{|l|l|l|l|l|l|}
    \hline
    \textbf{Rank}            & \textbf{Category} & \textbf{Output Type}   & \textbf{Total}   & \textbf{Percent}      & \textbf{Cumulative}    \\ \hline
    1                        & Text              & Plain                  & 977328           & 61.72                 & 61.72                  \\ \hline
    2                        & Image             & PNG                    & 339589           & 21.45                 & 83.17                  \\ \hline
    3                        & Text              & HTML                   & 208170           & 13.15                 & 96.32                  \\ \hline
    4                        & Application       & Javascript             & 22842            & 1.44                  & 97.76                  \\ \hline
    5                        & Application       & Jupyter JSON Widget    & 14532            & 0.91                  & 98.67                  \\ \hline
    6                        & Text              & Markdown               & 4750             & 0.31                  & 98.98                  \\ \hline
    7                        & Image             & SVG                    & 3854             & 0.24                  & 99.22                  \\ \hline
    8                        & Application       & Plotly v1 + JSON       & 3385             & 0.21                  & 99.45                  \\ \hline
    9                        & Text              & LaTeX                  & 2606             & 0.16                  & 99.61                  \\ \hline
    10                       & Text              & Plotly v1 + HTML       & 1623             & 0.10                  & 99.71                  \\ \hline
    11                       & Image             & JPEG/JPG               & 1326             & 0.08                  & 99.79                  \\ \hline
    12                       & Application       & BokehJS + JSON         & 902              & 0.06                  & 99.85                  \\ \hline
    13                       & Application       & PDF                    & 677              & 0.04                  & 99.89                  \\ \hline
    14                       & Application       & VDOM.v1 + JSON         & 330              & 0.02                  & 99.91                  \\ \hline
    15                       & Application       & BokehJS + JSON         & 326              & 0.02                  & 99.93                  \\ \hline
    16                       & Application       & Papermill + JSON       & 300              & 0.01                  & 99.94                  \\ \hline
    17                       & Application       & Holoviews Load + JSON  & 157              & 0.009                 & 99.95                  \\ \hline
    18                       & Application       & Holoviews Exec + JSON  & 122              & 0.007                 & 99.96                  \\ \hline
    19                       & Application       & JSON                   & 93               & 0.005                 & 99.96                  \\ \hline
    20                       & Application       & Colaboratory Intrinsic & 77               & 0.004                 & 99.97                  \\ \hline
    \multirow{3}{*}{21 - 36} & Application       & (12 Others)            & 445              & \multirow{3}{*}{0.03} & \multirow{3}{*}{100.0} \\ \cline{2-4}
                             & Image             & (1 Others - GIF)       & 19               &                       &                        \\ \cline{2-4}
                             & Text              & (1 Others - GraphVIZ)  & 10               &                       &                        \\ \hline
    \textbf{Total Outputs}   &                   &                        & \textbf{1583255} &                       & \textbf{100.0}         \\ \hline
    \end{tabular}
    \end{adjustbox}
    \label{appendix:output-types-list}
\end{table*}

\end{document}